\documentclass{jfm}
\usepackage{graphicx}
\usepackage{epstopdf, epsfig}
\usepackage{bm} 
\usepackage[normalem]{ulem}
\usepackage{xcolor} 
 \usepackage{amsmath} 
  \usepackage{array}
    \usepackage{eqnarray}
\newcolumntype{M}[1]{>{\centering\arraybackslash}m{#1}}

  \newcommand{\review}[1]{\textcolor{black}{#1}}
  
\newtheorem{lemma}{Lemma}
\newtheorem{corollary}{Corollary}

\title{On the origin of circular rolls in rotor-stator flow}

\author{
A. Gesla\aff{1,2}, Y. Duguet\aff{2}, P. Le Qu\'er\'e\aff{2} and L. Martin Witkowski\aff{3}
  \corresp{\email{laurent.martin-witkowski@univ-lyon1.fr}}
 }

\shorttitle{On the origin of circular rolls in rotor-stator flow}
\shortauthor{A. Gesla, Y. Duguet, P. Le Qu\'er\'e and  L. Martin Witkowski}

\affiliation{\aff{1}Sorbonne Universit\'e, F-75005 Paris, France
\aff{2}LISN-CNRS, Universit\'e Paris-Saclay, F-91400 Orsay, France
\aff{3}Universite Claude Bernard Lyon 1, CNRS, Ecole Centrale de Lyon, INSA Lyon, LMFA, UMR5509, 69622 Villeurbanne, France
}

\begin{document}

\maketitle

\begin{abstract}

Rotor-stator flows are known to exhibit instabilities in the form of circular and spiral rolls. 
While the spirals are known to emanate from a supercritical Hopf bifurcation, the origin of the circular rolls is still unclear. In the present work we suggest a quantitative scenario for the circular rolls as a response of the system to external forcing. We consider two types of axisymmetric forcing: bulk forcing (based on the resolvent analysis) and boundary forcing using direct numerical simulation. Using the singular value decomposition of the resolvent operator the optimal response is shown to take the form of circular rolls. 
The linear gain curve shows strong amplification at non-zero frequencies following a pseudo-resonance mechanism.
The optimal energy gain is found to \review{scale exponentially with the Reynolds number $Re$ (for $Re$} based on the rotation rate and interdisc spacing $H$). %\sout{, in connection with \review{strong} non-normality.}
The results for both types of forcing are compared with former experimental works and previous numerical studies. Our findings suggest that the circular rolls observed experimentally are the
% combined %review
effect of the high forcing gain \review{together with} the roll-like form of the leading response of the linearised operator. For high enough Reynolds number it is possible to delineate between linear and nonlinear response. For sufficiently strong forcing amplitudes, the nonlinear response is consistent with the self-sustained states found recently for the unforced problem. The onset of such non-trivial dynamics is shown to correspond in state space to a deterministic leaky attractor, as in other subcritical wall-bounded shear flows.
 %At the onset of 

%The linear and nonlinear responses are compared with the self-sustained states found recently for the unforced problem. \\

% This file contains instructions for authors planning to submit a paper to the {\it Journal of Fluid Mechanics}. These instructions were generated in {\LaTeX} using the JFM style, so the {\LaTeX} source file can be used as a template for submissions. The present paragraph appears in the \verb}abstract} environment. All papers should feature a single-paragraph abstract of no more than 250 words, which provides a summary of the main aims and results. 
\end{abstract}

\begin{keywords}
rotor-stator, axisymmetric rolls, harmonic forcing, noise-sustained oscillations, resolvent analysis, nonlinear states, transient dynamics
% Authors should not enter keywords on the manuscript, as these must be chosen by the author during the online submission process and will then be added during the typesetting process (see http://journals.cambridge.org/data/\linebreak[3]relatedlink/jfm-\linebreak[3]keywords.pdf for the full list)
\end{keywords}

% Dear Editor,

% Please find attached a new manuscript revisiting a classical subject : transition to unsteadiness in  rotor-stator flows, in particular the origin of axisymmetric waves often called circular rolls (CR) observed in experiments more than 30 years ago. The origin of the CRs has never been clearly explained and they do not correspond to an instability but are noise sustained structures.  This work includes theory, numerical simulations and detailed quantitative comparison to published experimental results.

% We believe that many of the results contained in this study can also have an impact, beyond the rotating flows community, on many fluid dynamicists interested in noise sustained structures, linear and nonlinear receptivity, and transition to turbulence.

% Yours sincerely,

%      Artur Gesla,  Yohann Duguet, Patrick Le Quéré and Laurent Martin Witkowski.

\section{Introduction}

The study of rotor-stator flows has a long history. Early investigations of the laminar flow regime in the flow between two infinite discs, one of which rotating, questioned the structure of the laminar velocity field. Many co-existing steady solutions with a self-similar spatial structure were reported \citep{holodniok1977computation,zandbergen1987karman}, among them the well-known solutions by \cite{Batchelor_qjmam_51} and \cite{Stewartson_pcps_53}. % \yd{\sout{The rotation of the disc induces a flow in the cavity formed.}}
In the presence of a radial shroud, the system  is thought to admit only one steady solution existing for all rotation rates, which will be considered as the \emph{base flow}. It features a closed meridional recirculation of the Batchelor type including a boundary layer on the stator and another one on the rotor.
%\lmw{Self similar solution has also one base flow solution. This statement, of course, depends on how we defined base flow solution. I would define base flow is when Navier Stokes admits a unique solution (in most cases for low enough Re values)}
This base flow solution departs increasingly from the self-similar solutions as the distance from the axis and the rotation rate 
% \agc{sure?, 0->100 lets say}
increase \citep{brady1987rotating}, which questions the quantitative relevance of all instability studies of self-similar solutions for the finite geometry.

We address here the mechanisms through which rotor-stator flows transition towards unsteady regimes interpreted as precursors of turbulent flow. Global linear stability analysis of the base flow \citep{Gelfgat_fdr_2015} predicts, for large enough aspect ratios, the linear instability of non-axisymmetric spiral modes, usually in quantitative agreement with concurrent experimental observations \citep{Gauthier_1998,Schouveiler_pof_1998} and later numerical simulations \citep{Serre_jfm_2001,Serre_pof_04}. The azimuthal wavenumber $m$ of these spirals is typically large and comparable to the radius-over-gap ratio $\Gamma$. The spiral modes and their onset were reported as experimentally robust, independent of the noise level \cite{Gauthier_1998}. The nonlinear saturation of these spiral modes follows a simple supercritical scenario followed by turbulent transition \citep{Launder_arfm_10}. 

The spiral arms are however not the coherent structures appearing at the lowest rotation rates. Concentric rolls of finite-amplitude have been frequently reported as the earliest manifestation of unsteadiness in moderate-to-large aspect ratios. This phenomenon was first reported in spin-down experiments \citep{Savas_jfm_1987} and then in most experimental studies \citep{Gauthier_jfm_1999,Schouveiler_jfm_2001}, but have been missed by most computational stability analyses. The presence of these additional flow structures can potentially undermine all predictions from linear stability analysis.  The concentric rolls were \review{first} reproduced in direct numerical simulations \review{following impulsive perturbations, but they were reported to convect towards the center and to vanish, being hence short transients} rather than sustained coherent structures \citep{Lopez_pof_2009}. \review{This is consistent {so far} with the convective instability viewpoint put forward in the recent review by \cite{martinand2023instabilities}, although such a viewpoint is intrinsically limited in finite geometries.}
% \lmw{ this sentence is coming a bit too early ? May be in next paragraph}. The observation of these axisymmetric structures motivated new \lmw{seems to relate to 2002 in the next sentence} investigations of both linear and nonlinear regimes of the rotor-stator flow in a strictly axisymmetric context.
As shown by \cite{Daube_cf_2002} and recently confirmed by \cite{gesla2023subcritical} in a strictly axisymmetric context, the base flow is linearly unstable to the axisymmetric mode for much higher Reynolds number $Re$ than suggested from experimental observation. 
% expected from the experimental results.
\review{This} agrees with previous linear stability studies that predict spirals as the first mode of linear instability. Chaotic subcritical solutions were actually identified by \cite{Daube_cf_2002}. However their continuation towards lower $Re$ has failed to reach the low values relevant for circular rolls in experiments \citep{gesla2023subcritical}. The riddle of the dynamical origin of the circular rolls reported in these experiments \review{thus} remains unsolved.

\review{Continuous efforts have been made in the past to link the stability of enclosed rotor-stator flows to the results of local stability analysis of one-dimensional similarity solutions. 
%There has been continuing efforts in the past to relate the stability of enclosed rotor-stator flows to results from local stability analysis of one dimensional similarity solutions. 
All these studies rely on boundary layer profiles and imply 
% either
unbounded domain{s} in the 
% axial and 
radial direction for one- or two-disc set-up{s}.
% or with infinite radial extent for two-discs. 
In the \review{latter} case, large enough Reynolds number\review{s are} also necessary so that the boundary layers are well separated. A recent review of transition scenarios in finite- and infinite-radius rotor-stator configurations based on a local stability analysis can be found in \cite{martinand2023instabilities}. The local stability analysis is necessarily
\review{approximative}
% approximate
for enclosed rotor-stator flow but nevertheless very useful to understand the physical mechanism of instabilities. %However, the scope of such analysis is restricted to in
%The local stability analysis is mostly used when analysing the self-similar velocity profiles in a geometry on infinite radial extent (either one-disc or two-disc set-ups). Despite some similarities,
% the transition mechanisms that will be investigated here appear fundamentally different from those reported in \cite{martinand2023instabilities}.
% such mechanisms are reported exclusively at higher Reynolds numbers, they feature non-axisymmetric modes, well separated boundary layers and display instability of the rotor boundary layer rather than the stator one. 
The Reynolds number range where the circular rolls first appear does \review{however} not correspond to separated boundary layers, which questions the quantitative relevance of local stability analysis. 
% In this work a global framework will be therefore preferred.
} 

A first hint \review{about the origin of the circular rolls} comes from the experimental study by \cite{Gauthier_jfm_1999}. They observed that a change of motor in their experimental set-up lowers the threshold of appearance of rolls by roughly half. This pointed in turn to a high sensitivity of the rolls to external disturbances. Following these observations, numerical computations were performed where the system was continuously forced with a sinusoidal libration of the rotor \citep{Lopez_pof_2009,do2010optimal}.
% \plq{I think we should discuss a little bit in more detail the different forcing ways, the first paper  by Lopez was restricted to $\omega = 1$ and $A = 0.05$, that by Do widened the range of parameters $\omega $ between 0.5 and 5, A between $10^{-6} $ and $0.05$ , but the use of optimal is somewhat misleading. We should also describe the kind of forcing used by Gauthier }\yd{Yes but we must also keep the intro compact and readable. This intro is not quantitative so far and the details you suggest to add are all very technical. Look rather at what is written in p. 17}
 This forcing also proved to sustain a roll-like response although the exact temporal dynamics would remain to be compared to its experimental counterpart. \review{Based on direct numerical simulation,} \cite{do2010optimal} demonstrated that nonlinear effects contributed to the global dynamics of the rolls. \review{We present results of the optimal forcing from the resolvent analysis that complement the direct numerical simulations and bring new insigths. Since the form of the forcing in \cite{do2010optimal} is fixed, their study can be understood as optimisation in the forcing frequency only. The form of the optimal response resulting from the resolvent analysis can be understood as a spatial optimisation and form a novelty with respect to \cite{do2010optimal}. } More recently it was demonstrated in \cite{gesla2023subcritical}, at least for the case $\Gamma=10$, that the axisymmetric system features, independently of any external forcing, a self-sustained nonlinear regime coexisting with the base flow below the critical threshold. The existence of this subcritical regime is reminiscent of other subcritical shear flows such as Poiseuille flows and plane Couette flow \citep{eckhardt2018transition,avila2023transition}. \review{This leads to the dilemma whether the circular rolls observed in experiments should be interpreted as a response to external forcing, in other words a noise-sustained state, or as the footprint of a self-sustained coherent state of nonlinear origin. Formulated differently, does the flow follow a resonance scenario or does it oscillate autonomously independently of the way it is forced? Even in the case where the flow is externally forced, is the response linear or are nonlinear interactions important? 
%  \sout{display its own preferred response at a given frequency, and if yes through which selection mechanism?}
 } 

%\textcolor{green}{Although the pieces of information were starting to point to the fact that the rolls observed experimentally could be noise-induced response, a few questions remained unanswered. Is the roll-like response and the frequency at which the system responds a preferred response of the system or is it just a result of the external forcing? Is the roll-like response a response to forcing or a footprint of a coherent nonlinear state? What causes the sudden appearance of the rolls at a sharp threshold in the experiment?}\\

The present study aims at answering these questions by means of resolvent analysis and numerical simulations. Although the circular rolls were reported experimentally only at low values of  $Re$ where the flow stays axisymmetric, we propose to consider $Re$ as the main governing parameter for the purely axisymmetric rotor-stator configuration, regardless of whether the experimental dynamics would stay axisymmetric or not, and to investigate the effect of varying $Re$ on both the linear receptivity scenario and the fully nonlinear one. \review{Due to the assumed axisymmetric configuration the present study does not bring any insight into optimally forced 3D structures.}
Our receptivity approach to external forcing follows several complementary approaches. Classical optimal linear response theory, through resolvent analysis, directly yields the optimal forcing eliciting the strongest response of the flow. By design however, this forcing is applied within the bulk of the flow, away from the solid walls. Parasite vibrations of the set-up are rather expected to act at the fluid-solid interface and, thus, bulk-based optimisation might not capture them especially when the forcing is modelled as an additive force. For this reason, as well as for the freedom of incorporating nonlinear fluid interactions, direct numerical simulation of fluid flow in the presence of well controlled boundary oscillations is also considered without any optimisation. As we shall see, the scenario most consistent with experimental observations is the boundary forcing.
%\ags{
%, although white noise and harmonic forcing do not come as the most evident forcing protocols in the frame of this comparison. Despite such evidence, the nonlinear subcritical scenario emerges as well in numerical forcing experiments \lmw{I am not sure to understand this sentence}. 
%}
%\agc{???}
For higher $Re$, we demonstrate numerically how circular rolls of finite amplitude can be elicited and sustained by nonlinear forcing, whereas depending on the value of $Re$ they correspond either to super-transients or to an attracting dynamics, should the forcing be turned off. \review{Beyond the immediate analogy with subcritical shear flows, it is also of strong
general interest as it yields a concrete example of nonlinear receptivity whereas receptivity is traditionally investigated using linear tools.}  

%\textcolor{red}{The ambiguity between the bulk forcing approach and the boundary forcing approach should be mentioned somewhere in this introduction, may be here} For a 2D system of moderate size, the resolvent can be found directly, and its singular values and singular vectors give the optimal gain, forcing, and response of the flow. A forced DNS can also yield the roll-like structure of the flow. Both tools are used extensively in the paper to \yd{show evidence that the origin of} the experimental observation is the \yd{response to the unsteady} forcing present in the experiment.

%Apart from explaining the circular rolls at low $Re$, the paper explores the regimes of higher $Re$ where the effects of nonnormality are severe, and the existence of self-sustained nonlinear states affects the overall dynamics. In particular, a parameter range where the system exhibits supertransient dynamics is characterized. \textcolor{red}{YD : let me imporve this part}

The structure of the paper is as follows. Section \ref{methods} introduces the governing equations and the discretized ones. Section \ref{base} is a description of the axisymmetric base flow. Section \ref{optfor} is devoted to finding optimal forcing using resolvent analysis. A comparison of the identified states with experimental evidence is presented in section \ref{expcomp}. The validity of the linear response assumption is verified in section \ref{nonlin} by considering both linearized and nonlinear time integration. Also, the amplitudes of forcing at which the nonlinearity plays an important role are characterized. The main findings of the paper are summarized in section \ref{summary}.

\begin{figure}
    \centering
    \includegraphics[width=0.5\columnwidth]{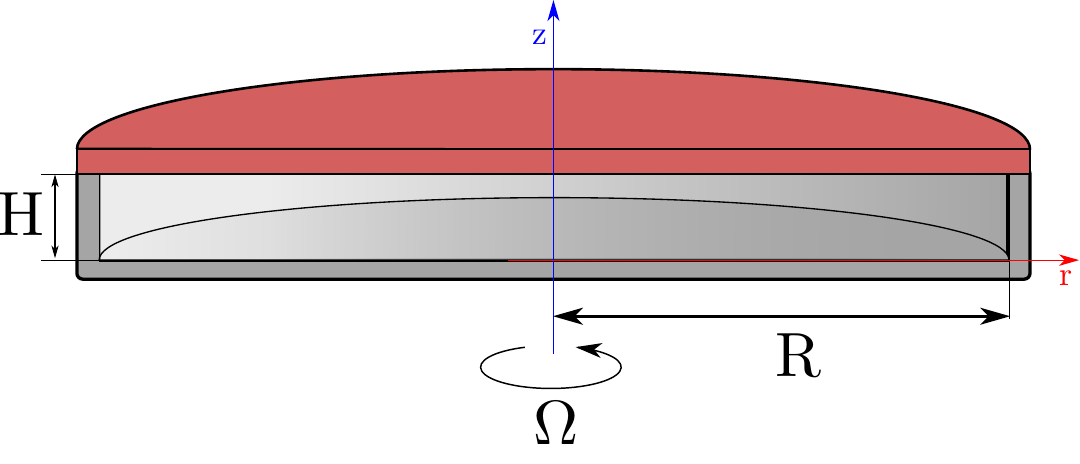}
 \caption{Rotor-stator geometry with rotating shroud. The set-up is characterised by the Reynolds number $Re= {\Omega H^2}/{\nu}$ and an aspect ratio $\Gamma=R/H$ fixed in most of this article to 10.}
 \label{fig:sketch}

\end{figure}

%\section{Methodology and numerical methods} \label{methods}
\section{Governing equations and numerical methods} \label{methods}

The system consists of two coaxial disks of  radius $R$, separated by a gap $H$. One of the disks, the rotor, rotates at a constant dimensional angular velocity $\Omega$, while the stator is at rest, see figure \ref{fig:sketch}. We chose to non-dimensionalise all lengths by the gap $H$ and time using $(\Omega)^{-1}$. Assuming a constant kinematic viscosity $\nu$ for the fluid, two non-dimensional parameters characterise this system, namely the geometric aspect ratio $\Gamma=R/H$ and the (gap-based) Reynolds number $Re=\Omega H^2/\nu$.
\review{Other possible definitions for the Reynolds number $Re$ include $Re_R=\Omega R^2/ \nu$, built on the length scale $R$ or $Re_{RH}={\Omega R H}/{\nu}$. In the current work the definition of $Re$ based on $H$ is preferred since it can be defined in both finite and infinite radial extent set-ups, as it facilitates a straightforward comparison with other studies.}

\subsection{Governing equations}

%The system is non-dimensionalised using the rotor's angular speed $\Omega_0$, the interdisk gap $H$ and the kinematic viscosity $\nu$. 
The non-dimensional velocity  $\mathbf{v}=(v_r,v_{\theta},v_z)$ and the non-dimensional pressure $\pi$ obey the incompressible Navier-Stokes equations (\ref{axiNSr}~--~\ref{axidiv}). Throughout the whole paper we assume that the flow is strictly axisymmetric. We consider  the Navier--Stokes equations in the cylindrical coordinate system $({\bm e}_r,{\bm e_{\theta}},{\bm e_{z}})$ :

\begin{equation} \label{axiNSr}
\frac{\partial v_r}{\partial t} + v_r \frac{\partial v_r}{\partial r} + v_z \frac{\partial v_r}{\partial z} -\frac{v^2_{\theta}}{r}= -\frac{\partial \pi}{\partial r} + \frac{1}{Re}(\frac{1}{r}\frac{\partial}{\partial r} (r\frac{\partial v_r}{\partial r}) - \frac{v_{r}}{r^2} + \frac{\partial^2 v_r}{\partial z^2})  
\end{equation}
\begin{equation} 
\label{axiNSt}
\frac{\partial v_{\theta}}{\partial t} + v_r \frac{\partial v_{\theta}}{\partial r} + v_z \frac{\partial v_{\theta}}{\partial z} +\frac{v_rv_{\theta}}{r}=  \frac{1}{Re}(\frac{1}{r}\frac{\partial}{\partial r} (r\frac{\partial v_{\theta}}{\partial r}) - \frac{v_{\theta}}{r^2} + \frac{\partial^2 v_{\theta}}{\partial z^2}) 
\end{equation}
\begin{equation} \label{axiNSz}
\frac{\partial v_z}{\partial t} + v_r \frac{\partial v_z}{\partial r} + v_z \frac{\partial v_z}{\partial z} = -\frac{\partial \pi}{\partial z} + \frac{1}{Re}(\frac{1}{r}\frac{\partial}{\partial r} (r\frac{\partial v_z}{\partial r}) + \frac{\partial^2 v_z}{\partial z^2})
\end{equation}
\begin{equation} \label{axidiv}
\frac{1}{r} \frac{\partial (r v_r)}{\partial r} + \frac{\partial v_z}{\partial z}=0,
\end{equation}

The coupled system of partial differential equations is complemented with the no-slip boundary conditions expressed  as \eqref{bcvecadim} :

%\begin{equation} \label{bc}
%\begin{cases}
%    u_{\theta}=\Omega r,& \text{on the rotor} \\
%    u_{\theta}=\Omega R,& \text{on the shroud} \\
%    u_{\theta}=0,& \text{on the stator}, \\
%  \end{cases}
%end{equation}

%\begin{equation} \label{bcvecdim}
%\begin{cases}
%    {\bm u}=\Omega r{\bm e_{\theta}},& \text{on the rotor} \\
%    {\bm u}=\Omega R{\bm e_{\theta}},& \text{on the shroud} \\
%    {\bm u}={\bm 0},& \text{on the stator}, \\
%  \end{cases}
%\end{equation}

\begin{equation} \label{bcvecadim}
\begin{cases}
    {\bm v}=r{\bm e_{\theta}},& \text{at the rotor $(z=0)$} \\
    {\bm v}=\Gamma {\bm e_{\theta}},& \text{at the shroud $(r=\Gamma)$} \\
    {\bm v}={\bm 0},& \text{on the stator $(z=1)$}, \\
  \end{cases}
\end{equation}

%\textcolor{red}{to write in non-dimensional form, here and everywhere}

%\begin{equation} \label{bcvec}
%\begin{cases}
%    {\bm u}=r{\bm e_{\theta}},& \text{on the rotor} \\
%    {\bm u}=\Gamma {\bm e_{\theta}},& \text{on the %shroud} \\
%    {\bm u}={\bm 0},& \text{on the stator}, \\
%  \end{cases}
%\end{equation}

%Reynolds number $Re_H=\Omega H^2/\nu$, $\Gamma=R/H$.
%\\
% \mathbf{\nabla} \cdot  \mathbf{u}&=0
%\end{equation}

\begin{table}
\centering
\begin{tabular}{lcccc}
resolution & $N_r$ & $N_z$ &type&DOF\\ 
R0  & 300&80&uniform& 99 k  \\
R1  & 600&160&uniform& 390 k\\
 R2 & 1024&192 &non-uniform&796 k \\
  R3 & 1536&288 &non-uniform&1.8 m \\
\end{tabular}
\caption{Mesh resolutions used in the study.}
\label{reolutions}
\end{table}

\subsection{Discretisation}
The continuous problem is discretised using a second order Finite Volume method on a staggered grid. The details of the discretisation as well as the staggered arrangement can be found in {appendix A of} \cite{Faugaret_jfm_2022}. Two types of mesh, uniform and non-uniform, are used. The non-uniform mesh is refined in the regions near the rotor, the stator and  the rotating shroud. Details on the nonuniform mesh setup can be found in \cite{gesla2023subcritical}. 
% Details on the resulting cell size for the pressure are given for two mesh types in Table \ref{reolutions}. In both uniform and non-uniform cases, the number of points in the radial direction $r$ (resp. the axial direction $z$) is noted $N_r$ (resp. $N_z$).

Details on the spatial discretisation used are given in Table \ref{reolutions}. In both uniform and non-uniform cases, the number of pressure cells in the radial direction $r$ (resp. the axial direction $z$) is noted $N_r$ (resp. $N_z$). For the application of the boundary conditions a layer of ghost cells is added outside the physical domain. 

%\agc{top is copied from prf}

   % \ag{+ mention the layer of ghost cells}

% In the radial direction it is uniform from $r=0$ to $r=8$ and then refined on the remaining $r\in(8,10)$ according to the formula 
% \begin{subequations} \label{meshr}
% \begin{align}
% % i=0,1,..,\frac{48}{70}nr\\
% % x_i=\frac{i\cdot (R-2)}{\frac{48}{70}nr} \\
% % i=\frac{48}{70}nr,...,nr\\
% % x_i=\frac{i\cdot 2}{\frac{22}{70}nr} !wrong\\
%     % u = 0.5*(1.+tanh(delta*(H*i/nz-0.5))/tanh(0.5*delta)) \\
%     % u = 0.5*(1.+tanh(delta*(H*i/nz-0.5))/)\\
%     b_i=\frac{1}{2}\left(1+\frac{tanh(\delta(x_i-\frac{1}{2}))}{tanh(\delta/2)}\right) \\
%     r_i=8+2\cdot\frac{b_i}{(a+(1-a) b_i)}    % u_i./(a+(1.-a)*u)\\
%     \end{align}
% \end{subequations}
% {with $\delta= 0.7258$ and $a=0.4989$, where $x_i\in(0,1)$ is the uniform mesh.} 70\% of total number of $N_r$ grid points are used in the uniform region $r\in(0,8)$ and the remaining 30\% are used in the non-uniform region $r\in(8,10)$.  The non-uniform mesh in the axial direction follows the formula
% \begin{subequations} \label{meshz}
% \begin{align}
% % i=0,1,..,nz\\
%     % u = 0.5*(1.+tanh(delta*(H*i/nz-0.5))/tanh(0.5*delta)) \\
%     % u = 0.5*(1.+tanh(delta*(H*i/nz-0.5))/)\\
%     z_i=\frac{1}{2}\left(1+\frac{tanh(\delta(\frac{i}{N_z}-\frac{1}{2}))}{tanh(\delta/2)}\right) 
%     % z=u./(a+(1.-a)*u)\\
%     \end{align}
% \end{subequations}
%    with $ \delta=2.8587 $. 

\subsection{Numerical methods}
% \agset 

% \agc{all this to rewrite}
 % Time integration is carried out using a Backwards Differentiation Formula 2 scheme (BDF2). It uses a prediction-projection algorithm in the rotational pressure correction formulation, as described in \cite{Guermond_cmame_2006}. The diffusion term treated implicitly gives rise to a Helmholtz problem for each velocity component increment. These Helmholtz problems are solved using an alternating-direction implicit (ADI) method in incremental form which preserves the second order accuracy in time. In the projection step the velocity field is projected onto the space of divergence-free fields by solving a Poisson equation for the pressure. This Poisson equation is solved using a direct sparse solver.
 % Time integration is performed with a time step $dt$ corresponding to a Courant–Friedrichs–Lewy (CFL) number of 0.3. }
 
%\yd{The system of Eq. \eqref{ns} admits for all $Re$ a steady solution $({\bm U}_b,P_b)$ labelled as \emph{base flow}. It is determined numerically using a Newton-Raphson algorithm.}

The nonlinear system of Eqs. (\ref{axiNSr}~--~\ref{axidiv}) admits for all $Re$ a steady  solution. Once the system is discretised, the solution of the large algebraic nonlinear system of equations is determined numerically using a Newton-Raphson algorithm (see e.g. appendix~B in~\cite{Faugaret_jfm_2022}). 
The $O(4N_rN_z)$ unknowns are the velocity and pressure values at each discretisation point. Due to the presence of the thin
boundary layers close to each disk, the set of equations resulting from the discretisation
of the continuous system is in general poorly conditioned. Sparse direct solvers will be
therefore preferred over the iterative solvers for solving the linear systems in each Newton-Raphson iteration.

A technical remark concerning the geometry can be made at this point : at the junction between the rotating shroud and the stationary disc, the sudden change in $u_{\theta}$ imposed in the boundary condition induces a discontinuity of the velocity field in the $(r,z)=(\Gamma,1)$ corner. This singularity can have an impact on the accuracy of the numerical solutions. In order to avoid singular boundary conditions and the well-known associated Gibbs phenomenon, several teams using pseudospectral solvers \citep{Serre_pof_04,Lopez_pof_2009} proposed to smooth out the discontinuous boundary condition between the stator and the shroud. Whenever Finite Volume discretisation is used the singular corner does not degrade the second order of spatial accuracy of the scheme, as demonstrated in \cite{gesla2023subcritical}. 

The linear stability of the base flow is evaluated using the Arnoldi method based on a well-validated ARPACK package \citep{Lehoucq_1998}. A generalised eigenvalue problem can be constructed using the Jacobian matrix used in the procedure of finding the base flow stemming from linearisation of the governing equations around a base flow (section \ref{optfor}).
For a generalised eigenvalue problem
\begin{equation}
    \bm{A} \bm{w}=\lambda \bm{B} \bm{w},
\end{equation}
the shift-and-invert method finds a subset of eigenvalues closest to a complex shift $s$ through repeated Arnoldi iteration:\begin{equation}
    \nu \bm{w}=\left( \bm{A}-s \bm{B} \right)^{-1}\bm{B}\bm{w}
\end{equation}
where the original eigenvalues $\lambda$ can be retrieved  with:
\begin{equation}
    \nu=\frac{1}{\lambda-s}
\end{equation}

The time integration of the governing equations  uses a prediction-projection algorithm in the rotational pressure correction formulation as described in \cite{Guermond_cmame_2006}. The prediction step combines a Backwards Differentiation Formula 2 (BDF2) scheme  for the diffusion terms and an explicit treatment of the convective terms, resulting in a Helmholtz problem for each velocity component increment. These Helmholtz problems are solved using an alternating-direction implicit (ADI) method in incremental form which preserves the second order accuracy in time. In the projection step the velocity field is projected onto the space of divergence-free fields by solving a Poisson equation for the pressure. This Poisson equation is solved using a direct sparse solver.
Once the base flow solution is found, the time-stepping code can be adapted to evolve the perturbation to the base flow rather than the full velocity field itself, at the expense of a triple evaluation of the convective terms. 
% \agunset

\section{Base flow} \label{base}
\begin{figure}
    \centering
    % \includegraphics{}
% \begin{table}
    % \centering
    % \begin{tabular}{M{0.1\textwidth}  M{0.49\textwidth} M{0.49\textwidth} }
    % & $u_{\theta}$ & $\psi$ \\ 
    % Re = 70 $\omega=0$ & \includegraphics[width=0.49\textwidth]{images/jfm-forcing70-0.png}
    % & \includegraphics[width=0.49\textwidth]{images/jfm-resp70-0.png} \\
    % Re = 150 $\omega=0$ & \includegraphics[width=0.49\textwidth]{images/jfm-forcing400-0.png}
    % & \includegraphics[width=0.49\textwidth]{images/jfm-resp400-0.png} \\
    % Re = 250 $\omega=0$ & \includegraphics[width=0.49\textwidth]{images/jfm-forcing1800-0.png}
    % & \includegraphics[width=0.49\textwidth]{images/jfm-resp1800-0.png} \\
    %  Re = 3000 $\omega=0$ & \includegraphics[width=0.49\textwidth]{images/jfm-forcing3000-0.png}
    % & \includegraphics[width=0.49\textwidth]{images/jfm-resp3000-0.png} \\
    % \end{tabular}
% \end{table}

\includegraphics[width=0.49\textwidth]{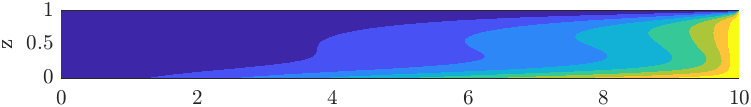}
\includegraphics[width=0.49\textwidth]{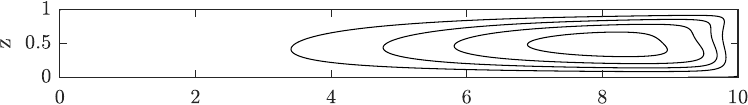}
\includegraphics[width=0.49\textwidth]{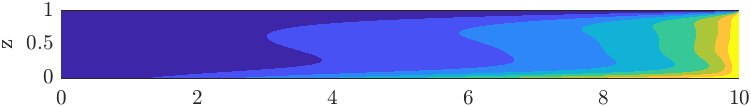}
\includegraphics[width=0.49\textwidth]{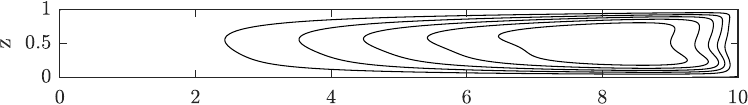}
\includegraphics[width=0.49\textwidth]{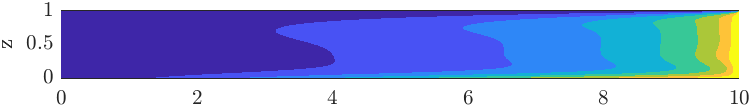}
\includegraphics[width=0.49\textwidth]{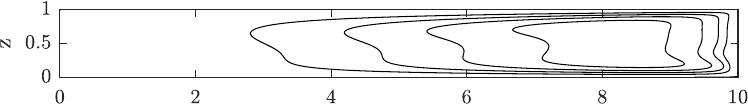}
\includegraphics[width=0.49\textwidth]{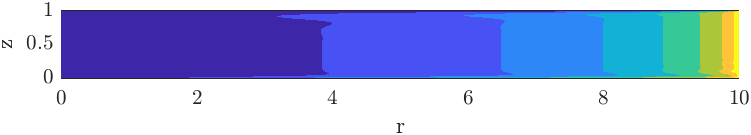}
\includegraphics[width=0.49\textwidth]{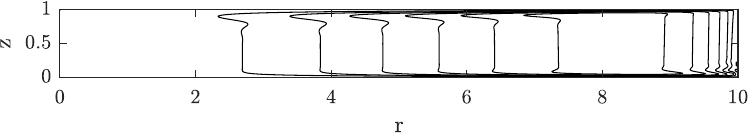}

    \caption{Axisymmetric base flow solution for $\Gamma=10$. Visualisations in a meridian section. From top to bottom : Re=70, 150, 250, 3000. Left: azimuthal velocity $u_{\theta}(r,z)$. The colormap divides the interval (0,10) in 8 equal subintervals.  Right: streamfunction $\psi(r,z)$. Plotted values : $\psi_{Re=70}=1,(1),4$. $\psi_{Re=150}=0.5,(0.5),2.5$. $\psi_{Re=250}=0.5,(0.5),2$. $\psi_{Re=3000}=0.1,(0.1),0.6$. $\psi=0$ corresponds to the wall.  Numerical resolution R1. 
    % \color{blue}{psi does not agree with PLQ 2001}
    }
    \label{base-evol}
\end{figure}

% \agc{notation}
% \subsection{2D base flow and its stability}
The base flow is the
% \ags{unique }
% \lmw{I would be less affirmative here} 
steady axisymmetric velocity field ${\bm U_b}$ solution to the governing equations (\ref{axiNSr}-\ref{axidiv}) compatible with the boundary conditions \eqref{bcvecadim}. It is associated with a pressure field $p_b$ defined up to an additive constant. It consists, for high enough $Re$, of two boundary layers, one on each disk, and a core region. Fluid is forced into circulation around the cavity by the rotation of the bottom disc. It is advected towards the rotating shroud and returns towards the axis along the stationary disk.  In figure \ref{base-evol}(left) isocontours of $u_{\theta}$ fields are plotted for  increasing $Re$. The meridional recirculation plane is shown using meridional streamlines in figure \ref{base-evol}(right). Streamfunction $\psi(r,z)$ is defined implicitly by $v_r=\frac{1}{r}\frac{\partial\psi}{\partial z}$ and $v_z=-\frac{1}{r}\frac{\partial\psi}{\partial r}$. For high enough $Re$ the core region between the boundary layers appears almost independent of $z$. 

% \begin{table}
%     \centering
%     \begin{tabular}{cccc}
%          reference & $Re_c$ & m & shroud \\
%              \citep{Daube_cf_2002}&2900--3000&  0 & rotating \\
%     \citep{Gelfgat_fdr_2015}&223.51& 19 & fixed\\
%     present & 223.90 & 19 & fixed\\
%     present & 528.91 & 32 & rotating\\

%     \end{tabular}
%     \caption{\review{Summary of linear stability analysis results for $\Gamma=10$.  To study the stability of the flow to 3D disturbances a modal ansatz in time and azimuthal direction is used \citep{Gelfgat_fdr_2015}. Critical $Re$ is listed along the number of modes $m$ in the azimuthal direction and a specification whether the shroud is rotating or not. Present results are in agreement with \citep{Gelfgat_fdr_2015} for the configuration with fixed shroud. TBC} }
%     %  
%     \label{tab:lasres}
% \end{table}
%  \review{Linear stability of the base flow to axisymmetric and non-axisymmetric disturbances can be investigated using modal ansatz in time and azimuthal direction \citep{Gelfgat_fdr_2015}. Linear instability threshold $Re_c$ for both axisymmetric and non-axisymmetric case is listed in table \ref{tab:lasres}.}.
 For the case $\Gamma=10$, a linear instability of the base flow in the axisymmetric configuration was found for $Re\approx3000$, due to the destabilisation of a wavepacket of counter-rotating circular rolls in the B\"odewadt layer at $r\approx2$ \citep{Daube_cf_2002}. This result was recently confirmed quantitatively by \cite{gesla2023subcritical}. Threshold for axisymmetric instability is one order of magnitude larger than the experimental threshold of $Re\approx200$ reported e.g. in \cite{Gauthier_1998}. 
 
 \review{The analysis of \cite{gesla2023subcritical} can be extended to the non-axisymmetric case by using an exponential \emph{ansatz} with an integer wavenumber $m$ in the azimuthal direction \citep{Gelfgat_fdr_2015}. The instability investigation of \cite{gesla2023subcritical} becomes spatially two-dimensional with an additional parameter $m$ (azimuthal wavenumber) and a critical value of $Re=528.91$ for $m=32$ (numerical resolution $N_r\times N_z=600\times128$ for $\Gamma=10$ set-up with rotating shroud). It is noted that the critical $Re$ for non-axisymmetric modes is around six times lower than for axisymmetric modes ($Re\approx3000$). The results for $m\neq0$ \review{concern} spiral rolls rather than circular roll. Since this work focuses only on circular rolls, it is assumed that the flow is strictly axisymmetric in what follows.}  
%\yd{This} is a first sign that the experimentally observed rolls are \yd{\emph{not}} the result of the linear instability. 

The base flow being linearly stable in the $Re$-interval of interest for this study, it can in principle be found by asymptotic time integration. Nevertheless, 
% \yd{\sout{for accuracy purposes and}}
in order to avoid strong transient growth effects \citep{Daube_cf_2002, gesla2023subcritical} it is safest to converge the base flow using the dedicated Newton-Raphson solver \review{to machine precision (i.e. when the Euclidean norm of the residual drops below $10^{-11}$). As shown in section \ref{nonlin}, for high enough $Re$ even a small perturbation will evolve to a top branch chaotic solution. Such a perturbation occurs for instance following a instantaneous change in $Re$. It was in particular found that the base flow cannot be found in a reasonable time using time integration for $Re>2100$. }

In the rest of the paper, once the base flow is known, we deal exclusively with axisymmetric perturbation ${\bm u}(r,z,t)$ to the base flow, i.e. ${\bm u}={\bm v}-{\bm U_b}$ and $p=\pi-p_b$, the equations for which can be deduced directly from the governing equations by subtraction. The associated boundary conditions are homogeneous of Dirichlet type, i.e. ${\bm u}=0$ at all walls. At the axis ($r=0$) $u_r$ and $u_{\theta}$ vanish while $u_z$ verifies a Neumann condition. %\agc{uz on axis is hom Neumann}

%\ag{+ specify somewhere that perturbation means always field - base}

\section{Linear response to forcing} \label{optfor}

%To analyse a forcing scenario possibly present in the experimental setup a forcing present only on the domain boundary is additionally analysed. This could be interpreted as inaccuracies in the motor rotation rate or the parasite vibrations that are the source of the perturbation at the system boundary. By changing the rotation rate of the rotor in a random manner we show that flow response is in the shape of circular rolls convected towards the axis. When a forcing signal, consisting only of harmonics of disc rotation rate,  is imposed, we achieve a perfect agreement with the experimental observation. 

\subsection{Optimal response theory} \label{optcomp}
% \agc{lets change the perturbation notation to small u, without a prime. Base is in capital anyway. This way notation will be clearer. I think adding a prime to almost every figure is a mistake. }\yd{YD : OK, then we need to have another notation for the full velocity/pressure. I suggest $v=U+u$ and $\pi=P+p$. Done. Let me know if I have forgotten to modify it somewhere}
In this subsection the linear response of the flow to the forcing is analysed using optimal response theory. Following \cite{cerqueira2014eigenvalue} we conduct the input/output analysis and show that the optimal response of the flow is for most relevant values of $Re$ in the shape of circular rolls, and that associated with high levels of optimal gain. The nonlinear system of equations (\ref{axiNSr}~--~\ref{axidiv}) can be linearised around the base flow when perturbation velocities are small enough.
If a forcing field ${\bm f}$ is introduced,
the resulting system for the perturbation field $({\bm u},p)$ can be rewritten 
\begin{eqnarray} 
\label{Nspert} 
\frac{\partial {\bm u}}{\partial t} + ({\bm U_b}\cdot {\bm \nabla}){\bm u}+ ({\bm u}\cdot {\bm \nabla}){\bm U_b}=-{\bm \nabla}p + \frac{1}{Re}{\bm \nabla}^2 {\bm u} + {\bm f} \: , \\ 
\label{Nspertd} {\bm \nabla} \cdot {\bm u}=0\: .
\end{eqnarray}
%which can be symbolically recast into the more compact form
%\begin{eqnarray} 
%\label{Nspert2} 
%\frac{\partial {\bm u'}}{\partial t} = \mathbf{L}{\bm u'} + {\bm f}, \\ 
%0=\mathbf{L_2} {\bm u'}. 
%\end{eqnarray}
The coupled system in \eqref{Nspert}-\eqref{Nspertd} is linear in ${\bm u}$, $p$ and also ${\bm f}$, which suggests to use the resolvent formalism. The theory by \cite{trefethen2020spectra} forms the ideal framework except that the original linear system needs be rewritten in a form 
% $\partial {\bm z}/\partial t={\bm L\bm z} + {\bm f_z}(t)$, where the unknown field ${\bm z}$,  a linear operator ${\bm L}$ and a forcing ${\bm f_z}(t)$
\review{$\partial {\bm q}/\partial t={\bm L\bm q} + {\bm f_q}(t)$, where the unknown field ${\bm q}$,  a linear operator ${\bm L}$ and a forcing ${\bm f_q}(t)$}
need to be specified.
% Following  \citep{cerqueira2014eigenvalue}, 
We introduce
the new variable $\mathbf{q}=(u_r,u_{\theta},u_z,p)$ which contains the values of the fields $u_r$,$u_{\theta}$, $u_z$ and $p$ at all the
%\ags{ interior}
% \lmw{why specify interior while we know it is not true. Or it is a shortcut beacuse it is mandatory to avoid BC points and that P matrix takes care of it ?}
points of the discretised domain. The size of $\mathbf{q}$ is $O(4N_rN_z)$. We also introduce the rectangular prolongation operator ${\bm P}$ of size $O(4N_rN_z \times 3N_rN_z)$ which maps ${\bm u}$ into ${\bm q}$, so that $\mathbf{P}^{T}\mathbf{q}=\mathbf{u}$ with the property $\mathbf{P}^{T}\mathbf{P}=\mathbf{I}$, see e.g. \cite{jin2021energy}. The linear system (\ref{Nspert}-- \ref{Nspertd}) can then be rewritten into the new form
\begin{equation}
\mathbf{B}\frac{\partial \mathbf{q}}{\partial t}=\mathbf{A}\mathbf{q}+\mathbf{P}{\mathbf{f}},
\end{equation}
% \ag{+addsizes}
where $\mathbf{B}=\mathbf{P}\mathbf{P}^{T}$. After a Fourier transform in time, each Fourier component $\hat{\mathbf{q}}$ of $\mathbf{q}$ satisfies
\begin{equation} \label{ioba}
    \left( i \omega \mathbf{B}-\mathbf{A} \right) \hat{\mathbf{q}}=\mathbf{P}\hat{\mathbf{f}}   \: ,
\end{equation}
resulting in the Fourier components $\mathbf{\hat{u}}$ of the velocity field $\mathbf{u}$ as the action of a matrix $\mathbf{R}$ on the forcing $\mathbf{\hat{f}}$ :
\begin{equation} \label{ressol}
    \hat{\mathbf{u}}=\mathbf{R}\hat{\mathbf{f}}
\end{equation}
where 
\begin{equation} \label{eq:res}
    \mathbf{R}=\mathbf{P}^{T}\left( i \omega \mathbf{B}-\mathbf{A} \right)^{-1}\mathbf{P}\\
\end{equation}
is the resolvent operator associated with the (real) angular frequency $\omega$.\\

% Afun3=@(x) P2'*((sigma*B-A)'\((P2*P2')*((sigma*B-A)\(P2*x))));
% tic; [evc,evs]=eigs(Afun3,[l22],1,'largestabs',IsFunctionSymmetric=true);toc; 

An optimal gain can be evaluated by identifying an optimal forcing for a suitable  norm of the resolvent $\mathbf{R}$. A positive symmetric linear operator $\mathbf{Q} \neq \mathbf{I}$, associated with the cylindrical coordinate system, can be used to define the  inner  product 
\begin{equation}
    <\mathbf{u},\mathbf{v}>_{\mathbf{Q}}=\mathbf{u}^*\mathbf{Q}\mathbf{v}=\int_0^1 \int_0^{\Gamma} (u_r^*v_r+u_{\theta}^*v_{\theta}+u_z^*v_z)r dr dz,
\end{equation}
where the asterisk denotes complex conjugate. The associated vector norm is defined by
% The norm of the resolvent $\mathbf{R}$ induced by the vector norm:
\begin{equation}
    \vert \vert \mathbf{u} \vert \vert_{\mathbf{Q}} =\sqrt{\mathbf{u}^*\mathbf{Z}^{T}\mathbf{Z}\mathbf{u}}=\vert \vert \mathbf{Z}\mathbf{u} \vert \vert_2,
\end{equation}
where $\mathbf{Q}=\mathbf{Z}^{T}\mathbf{Z}$. $\mathbf{Q}$ can be also used to define the following norm for the resolvent operator
\begin{equation}
    \vert \vert \mathbf{R} \vert \vert_{\mathbf{Q}}=\vert \vert \mathbf{Z}\mathbf{R}\mathbf{Z}^{-1} \vert \vert_2=\sigma_{1} (\mathbf{Z}\mathbf{R}\mathbf{Z}^{-1}),
\end{equation}
where $\sigma_{1}$ denotes the largest singular value in the SVD decomposition. 
% \plq{reference needed, Cerqueira, Trefethen ?}
 Note that finding the largest singular value of $\mathbf{ZRZ^{-1}}$ is equivalent to finding the largest eigenvalue of the eigenvalue problem \citep{cerqueira2014eigenvalue}:

\begin{equation} \label{evp}
    \mathbf{R}^*\mathbf{Q}\mathbf{R}\mathbf{\hat{f}}=\lambda\mathbf{Q}\mathbf{\hat{f}}
\end{equation}

where 
\begin{equation} \label{gain}
G=\lambda=\sigma_{1}^2=\frac{\mathbf{\hat{u}}^*\mathbf{Q}\mathbf{\hat{u}}}{\mathbf{\hat{f}}^*\mathbf{Q}\mathbf{\hat{f}}}
\end{equation} 
is interpreted as the optimal energy gain. 
% \agc{check this with sipp}
%In what follows the generalised Hermitian eigenvalue problem \eqref{evp} will be solved using a standard sparse SVD algorithm \yd{(REF NEEDED?)} to find the optimal gain. 

%The matrix $\mathbf{Q}$ differs from the identity matrix because of the .
%where the diagonal matrix $\mathbf{Q}=\mathbf{L}^{T}\mathbf{L}$ 
% \textcolor{red}{
%s the kernel of the scalar product

% \sout{corresponds to the scalar product in the} 

% }
%and:
%\begin{equation} \label{ressol}
%    \left( i \omega \mathbf{B}-\mathbf{A} \right)^{-1}\mathbf{P}\hat{\mathbf{f}}=\hat{\mathbf{q}}
%\end{equation}
%with the rectangular prolongation matrix $\mathbf{P}$ such that:
%$\mathbf{P}^{T}\mathbf{P}=\mathbf{I}$ and $\mathbf{P}\mathbf{P}^{T}=\mathbf{B}$.
% \textcolor{red}{Please define ${\bf P}$ at this stage.}
%By restricting the $\mathbf{\hat{q}}$ to only the velocity components \textcolor{red}{that is not clear} \ag{q is multiplied by Pt to get u, it is also not so clear in Sipp, I had to dig a bit to understand that P is rectangular}:
%\begin{equation} \label{ressol}
%    \mathbf{P}^T\left( i \omega \mathbf{B}-\mathbf{A} \right)^{-1}\mathbf{P}\hat{\mathbf{f}}=\hat{\mathbf{u}}
%\end{equation}
%a resolvent can be found:
% \textcolor{red}{YD : please re-read Cerqueira \& Sipp, they do not start exactly from the classical form of the Navier-Stokes equations to get tto the next formula. They also use a mass matrix $M$ by the way, do you understand well what it is?}
% Resolvent is defined:

%\textcolor{red}{This is a norm rather than a scalar product}

\subsection{Optimal response : results}

\begin{figure} 
    \centering
        \includegraphics[width=0.49\textwidth]{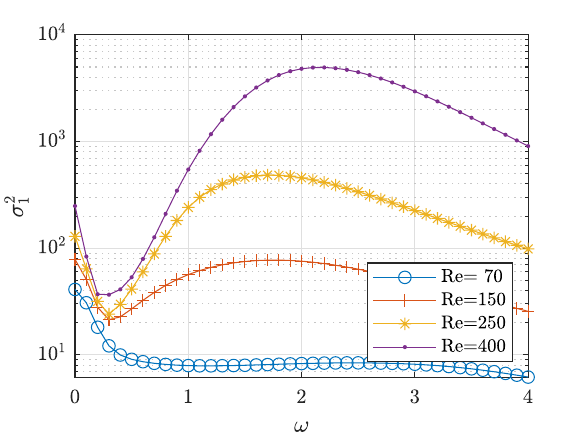}
                \includegraphics[width=0.49\textwidth]{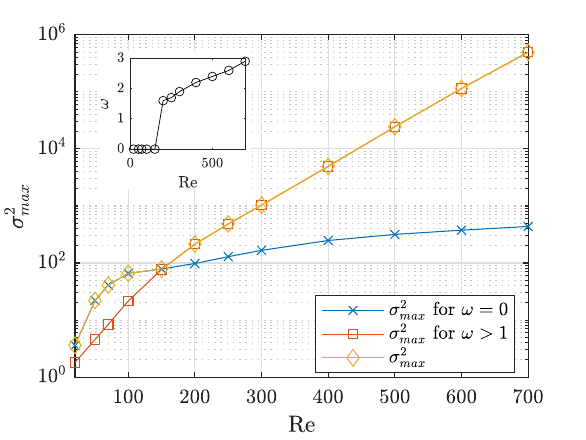}
    \caption{Optimal response for bulk-based forcing. Left: optimal energy gain as a function of forcing angular frequency $\omega$, given by the largest singular value of the resolvent operator \eqref{eq:res}. Right: optimal gain across all $\omega$'s as a function of $Re$. Numerical resolution R0. \review{The inset shows the optimal forcing angular frequency as listed in table \ref{taboptgain}}.}
    \label{optgain}
\end{figure}

For a range of angular frequencies $\omega$ the eigenvalue problem \eqref{evp} is solved numerically in MATLAB with the \textit{eigs()} function. 
% The resolvent is never computed explicitly but rather an action of the left hand side of \eqref{evp} is defined as a separate function and passed to \textit{eigs()} \lmw{it is nice details but a bit obscure.  Do you want to say that you are doing a LU decomposition of sparse (i*omega*B - A) ? }. 
The resulting eigenvalue $G=\sigma_1^2$ from Eq. \ref{gain} is plotted in figure \ref{optgain} (left) as a function of the (real) angular frequency of the forcing. The optimal value $\sigma_{max}^2$ over all angular frequencies obtained for various values of $Re$ are compared in figure \ref{optgain}(right), with the case $\omega=0$ singled out. The optimal value of $G$ gain over all $\omega$'s, i.e. the optimal gain, is also listed in the table \ref{taboptgain} together with the values obtained with different mesh resolutions.\\

% \textcolor{red}{Here, roughlym we have the choice between applying the SVD routine to $R$ or finding the spectrum of $R^*R$. Which choice should we focus on?}

\begin{table}
    \centering
    \begin{tabular}{ccccccc}
    Re & $\sigma_{max}^2$(R0)& $\sigma_{max}^2$(R1)&$\sigma_{max}^2$(R2)&$\sigma_{max}^2$(R3)& $\omega$ \\
20   & $3.61\times10^{0}$ &   &   &   & 0.0 \\
50   & $2.20\times10^{1}$ &   &   &   & 0.0 \\
70   & $4.10\times10^{1}$ & $4.10\times10^{1}$  &   &   & 0.0 \\
100  & $6.59\times10^{1}$ &   &   &   & 0.0 \\
150  & $7.85\times10^{1}$ & $7.86\times10^{1}$  & $7.87\times10^{1}$  &   & 0.0 \\
200  & $2.14\times10^{2}$ &   &   &   & 1.6 \\
250  & $4.84\times10^{2}$ & $4.94\times10^{2}$  & $4.90\times10^{2}$  &   & 1.7 \\
300  & $1.04\times10^{3}$ &   &   &   & 1.9 \\
400  & $4.95\times10^{3}$ & $5.50\times10^{3}$  & $5.47\times10^{3}$  &   & 2.2 \\
500  & $2.45\times10^{4}$ &   &   &   & 2.4 \\
600  & $1.15\times10^{5}$ &   &   &   & 2.6 \\
700  & $4.97\times10^{5}$ & $7.62\times10^{5}$  & $7.96\times10^{5}$  & $8.38\times10^{5}$  & 2.9 \\
1800 &  & $7.10\times10^{11}$ & $1.29\times10^{12}$ & $1.50\times10^{12}$ & 4.3 \\
3000 &  & $2.08\times10^{16}$ & $1.28\times10^{17}$ & $1.66\times10^{17}$ & 5.4\\
\end{tabular}
\label{taboptgain}
    \caption{Optimal forcing gain. The last column shows the optimal angular frequency $\omega$ associated with the largest energy gain (the computations are performed with a step of 0.1 in angular frequency). Four mesh resolutions (see table \ref{reolutions}) are used to find the optimal gain value. While for $Re\approx250$ results  can be considered as satisfactory for resolution R1, increased resolution is needed for larger $Re$. This is due to increasingly thin boundary layers required to resolve the optimal forcing mode. 
    % and  corresponds to the results on the finest mesh used. 
    }
\end{table}

%-------first the omega=0 response
\begin{figure}
    \centering
   
% \begin{table}
    \centering
    \begin{tabular}{M{0.15\textwidth}  M{0.42\textwidth} M{0.42\textwidth} }
    & forcing & response \\ 
    Re = 70 $\omega=0$ & \includegraphics[width=0.4\textwidth]{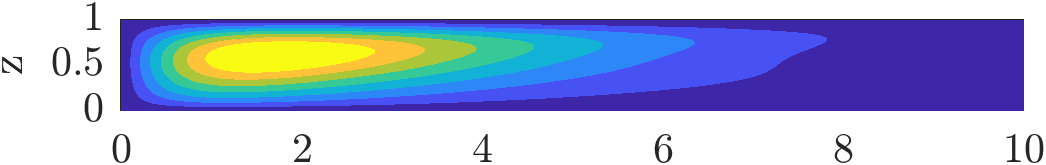}
    & \includegraphics[width=0.4\textwidth]{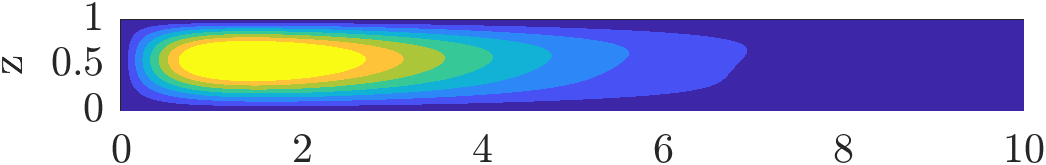} \\
    % Re = 250  $\omega=1.7$& \includegraphics[width=0.4\textwidth]{images/jfm-forcing250-1.7.png} & \includegraphics[width=0.4\textwidth]{images/jfm-resp250-1.7.png} \\
    Re = 150  $\omega=0$& \includegraphics[width=0.4\textwidth]{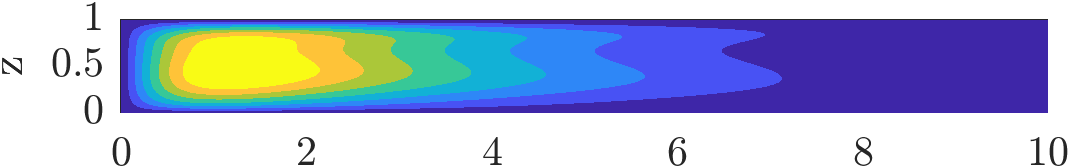} & \includegraphics[width=0.4\textwidth]{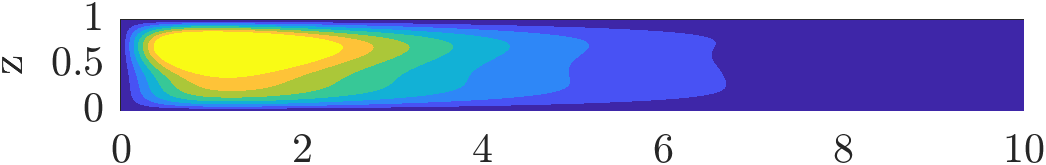} \\
    % Re = 150  $\omega=1.8$& \includegraphics[width=0.4\textwidth]{images/jfm-forcing150-1.8.png} & \includegraphics[width=0.4\textwidth]{images/jfm-resp150-1.8.png} \\
    % Re = 250  $\omega=1.7$& \includegraphics[width=0.4\textwidth]{images/jfm-forcing250-1.7.png} & \includegraphics[width=0.4\textwidth]{images/jfm-resp250-1.7.png} \\
         Re = 1800 $\omega=0$ & \includegraphics[width=0.4\textwidth]{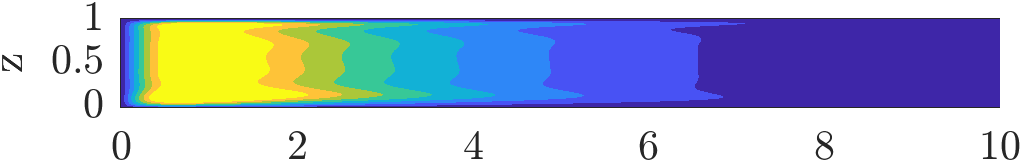}
    & \includegraphics[width=0.4\textwidth]{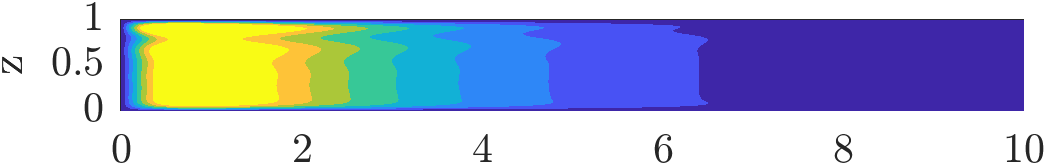} \\
     Re = 3000 $\omega=0$ & \includegraphics[width=0.4\textwidth]{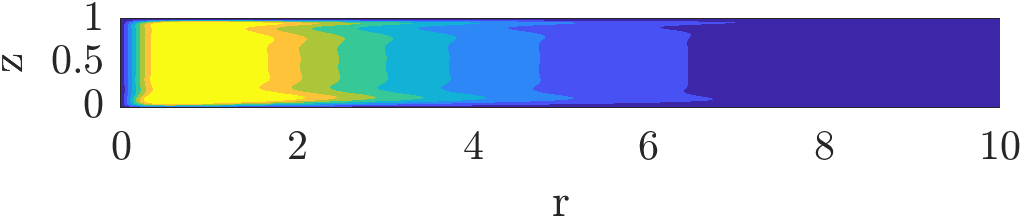}
    & \includegraphics[width=0.4\textwidth]{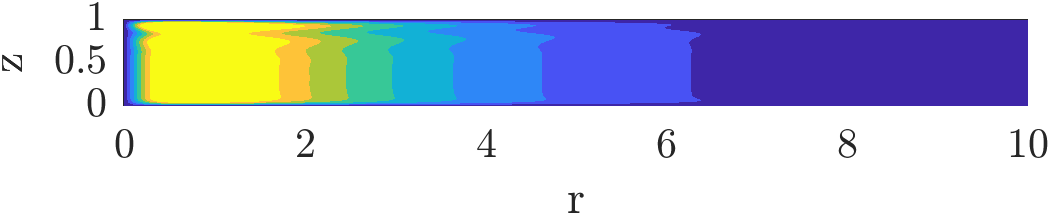} \\
    % Re = 3000 $\omega=2$& \includegraphics[width=0.4\textwidth]{images/jfm-forcing3000-2.png}
    % & \includegraphics[width=0.4\textwidth]{images/jfm-resp3000-2.png} 
    \end{tabular}
    \caption{Azimuthal perturbation velocity $u_{\theta}(r,z)$ for optimal forcing and optimal response for vanishing forcing frequency $\omega=0$.  For Re=70 and 150 the optimal forcing at $\omega=0$ is also optimal across all $\omega$'s. The colormap spans 8 equal subintervals of $[0,0.5]$. Both optimal forcing and response are normalised such that $\vert \vert \mathbf{\hat{f}} \vert \vert_{\mathbf{Q}}=\vert \vert \mathbf{\hat{u}} \vert \vert_{\mathbf{Q}}=1$.}
% \end{table}
 % \includegraphics{}
    \label{om0forcing}
\end{figure}

For low $Re<150$ the optimally amplified frequency is always $\omega=0$ . This corresponds to a steady forcing $f$ in the equations (\ref{axiNSr}~--~\ref{axidiv}). Forcing a given flow at $\omega=0$ can be interpreted in different complementary ways. It can first be understood as the smooth $\omega \rightarrow 0$ limit of a given harmonic forcing at frequency $\omega$. It can also be understood, in the unforced problem, as the steady streaming component associated with the nonlinear self-interaction of arbitrary oscillatory perturbations (see e.g. \cite{mantic_prl_2014}). In both cases the optimal response at $\omega=0$ is interpreted as an optimal steady mean flow correction, which justifies the special focus on $\omega=0$. 
% \plq{je ne suis pas sur de bien comprendre l'argument} \ag{!!!} 
As seen in figure \ref{om0forcing} the most amplified steady forcing is always localised in the region near the axis. For $Re>1800$ it inherits the characteristics of the base flow in the sense that it is composed of two boundary layers and invariant core in between.\\

\begin{figure}
    \centering
   
% \begin{table}
    \centering
    \begin{tabular}{M{0.05\textwidth}  M{0.47\textwidth} M{0.47\textwidth} }
    & forcing & response \\ 
    % $p$ Re = 250  $\omega=1.7$& \includegraphics[width=0.4\textwidth]{images/jfm-forcing250-1.7-p.png} & \includegraphics[width=0.4\textwidth]{images/jfm-resp250-1.7-p.png} \\
    $u_{r}$& \includegraphics[width=0.45\textwidth]{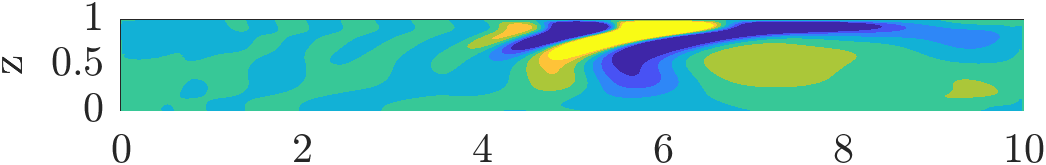} & \includegraphics[width=0.45\textwidth]{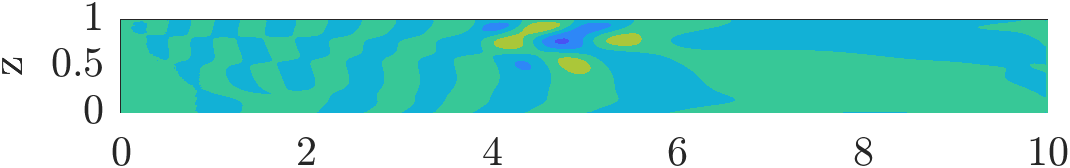} \\
    $u_{\theta}$& \includegraphics[width=0.45\textwidth]{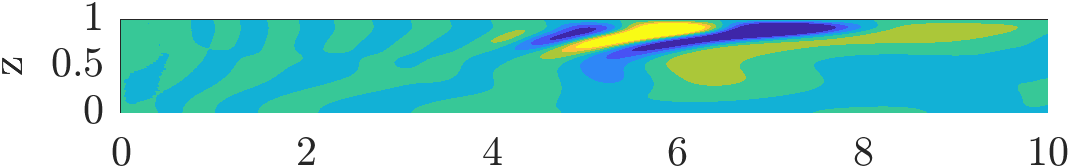} & \includegraphics[width=0.45\textwidth]{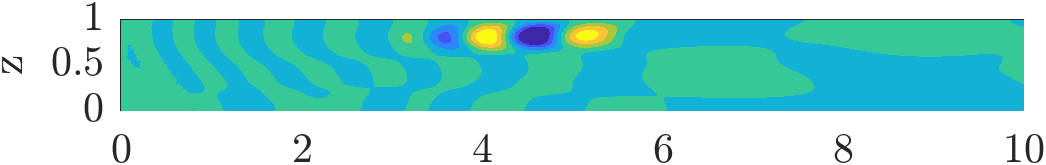} \\
    $u_{z}$& \includegraphics[width=0.45\textwidth]{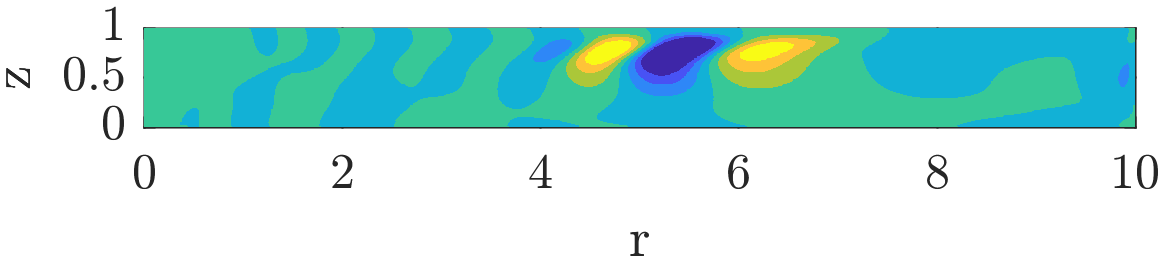} & \includegraphics[width=0.45\textwidth]{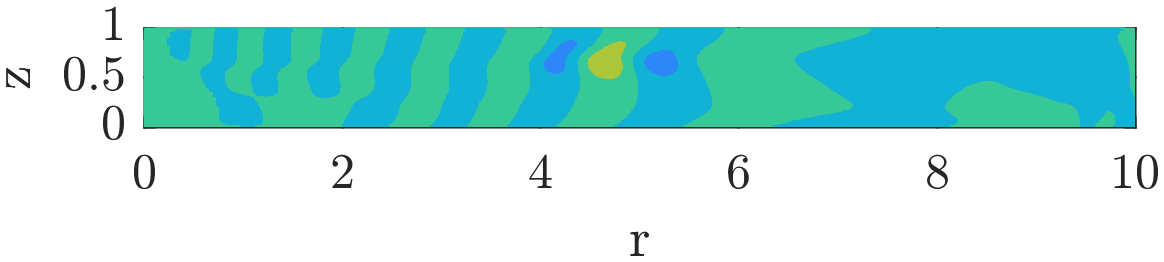} \\
    \end{tabular}

    % \begin{tabular}{M{0.15\textwidth}  M{0.42\textwidth} M{0.42\textwidth} }
    % & forcing & response \\ 
    % $p$ Re = 250  $\omega=1.7$& \includegraphics[width=0.4\textwidth]{images/jfm-forcing250-1.7-p-2.png} & \includegraphics[width=0.4\textwidth]{images/jfm-resp250-1.7-p-2.png} \\
    % $u_{r}$ Re = 250  $\omega=1.7$& \includegraphics[width=0.4\textwidth]{images/jfm-forcing250-1.7-u-2.png} & \includegraphics[width=0.4\textwidth]{images/jfm-resp250-1.7-u-2.png} \\
    % $u_{\theta}$ Re = 250  $\omega=1.7$& \includegraphics[width=0.4\textwidth]{images/jfm-forcing250-1.7-v-2.png} & \includegraphics[width=0.4\textwidth]{images/jfm-resp250-1.7-v-2.png} \\
    % $u_{z}$ Re = 250  $\omega=1.7$& \includegraphics[width=0.4\textwidth]{images/jfm-forcing250-1.7-w-2.png} & \includegraphics[width=0.4\textwidth]{images/jfm-resp250-1.7-w-2.png} \\
    % \end{tabular}
    
 \caption{Optimal forcing and response for Re=250. From top to bottom : perturbation velocity components $u_r$, $u_{\theta}$ and $u_z$ for the optimal angular frequency $\omega=1.7$.  The colormap spans 8 equal subintervals of (-0.15 0.15) for the optimal forcing, and (-0.6 0.6) for the optimal response plots. The optimal forcing and response are normalised such that $\vert \vert \mathbf{\hat{u}} \vert \vert_{\mathbf{Q}}=\vert \vert \mathbf{\hat{f}} \vert \vert_{\mathbf{Q}}=1$.%\agc{add r bot fig}
 }% \end{table}
    \label{fig:4components}
\end{figure}

For $Re\approx150$ the optimally forced structures change radically as unsteady ($\omega \neq 0$) forcing takes over steady forcing. This is clearly seen in figure \ref{optgain}(right) where for $Re \gtrsim 150$ nonzero forcing frequencies $\omega$ start to dominate the optimal gain curve. The structure of the associated optimal forcing lies entirely within the B\"odewadt layer, see figure \ref{fig:4components} which focuses on $\omega \approx 1.7$ close to the optimal angular frequency (see Table \ref{taboptgain}). For all velocity components these structures are tilted by the mean shear into the streamwise direction.
% \yd{I removed the analogy with other shear flows like Poiseuille because their optimal forcing requency is usually zero}.
While the amplitude of optimal forcing is similar in the three spatial components the corresponding optimal response may differ from component to component \citep{jovanovic2005componentwise}, and in the present geometry it is clearly dominated by the azimuthal component.  The signature of the circular rolls can be also seen in the azimuthal response although the corresponding structures should more realistically be labelled as azimuthal streaks. Their position ($r\approx 4-6$) 
% and the corresponding $Re$
is perfectly consistent with the experimental observations \review{by \cite{Schouveiler_jfm_2001} and the numerical studies of \cite{Lopez_pof_2009} and \cite{do2010optimal}}. Most importantly, and this constitutes one of the main findings of the current work, the optimal response is in the shape of circular rolls in the B\"odewadt layer.\\

%Before any more is said about the linear or nonlinear nature of the true mechanics responsible for the experimental observation, 

%This could explain the general agreement of the experimental studies that observe the circular rolls at $Re\approx200$. 
%It is also probably this steep increase of the optimal gain that could led authors to believe that the rolls are originating from a Hopf bifurcation at a fixed $Re$ \cite{Gauthier_jfm_1999}. 

\begin{figure}
    \centering
   
% \begin{table}
    \centering
    \begin{tabular}{M{0.15\textwidth}  M{0.42\textwidth} M{0.42\textwidth} }
    & forcing & response \\ 
    % Re = 70 $\omega=0$ & \includegraphics[width=0.4\textwidth]{images/jfm-forcing70-0.png}
    % & \includegraphics[width=0.4\textwidth]{images/jfm-resp70-0.png} \\
    % Re = 250  $\omega=1.7$& \includegraphics[width=0.4\textwidth]{images/jfm-forcing250-1.7.png} & \includegraphics[width=0.4\textwidth]{images/jfm-resp250-1.7.png} \\
    % Re = 150  $\omega=0$& \includegraphics[width=0.4\textwidth]{images/jfm-forcing150-0.png} & \includegraphics[width=0.4\textwidth]{images/jfm-resp150-0.png} \\
    Re = 150  $\omega=1.8$& \includegraphics[width=0.4\textwidth]{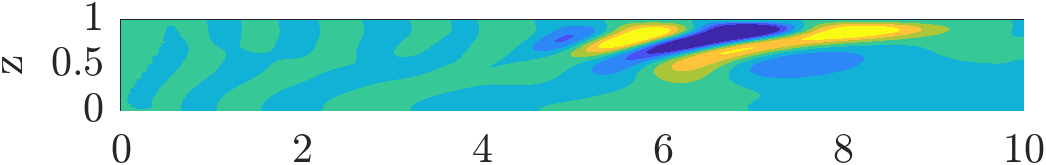} & \includegraphics[width=0.4\textwidth]{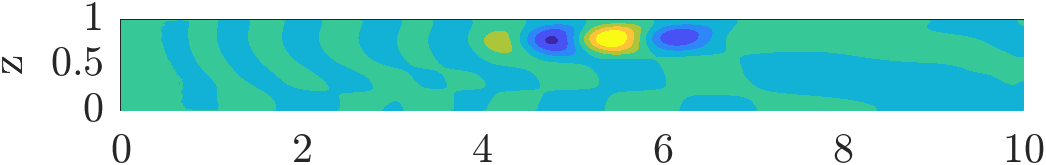} \\
    Re = 250  $\omega=1.7$& \includegraphics[width=0.4\textwidth]{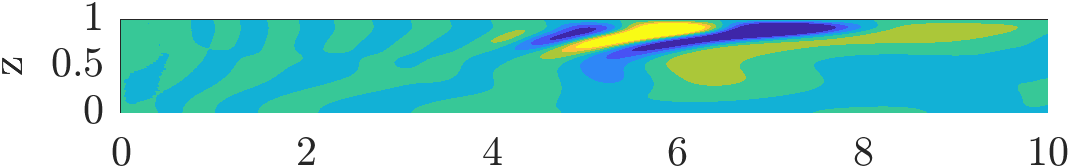} & \includegraphics[width=0.4\textwidth]{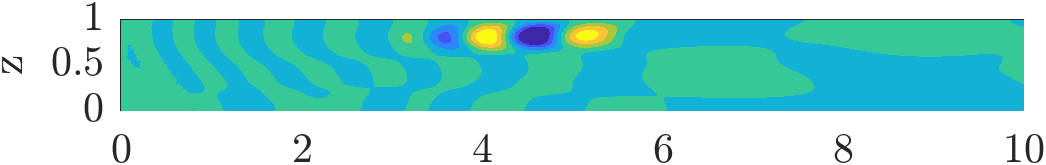} \\
    %  Re = 3000 $\omega=0$ & \includegraphics[width=0.4\textwidth]{images/jfm-forcing3000-0.png}
    % & \includegraphics[width=0.4\textwidth]{images/jfm-resp3000-0.png} \\
    % Re = 400 $\omega=2$& \includegraphics[width=0.4\textwidth]{images/jfm-forcing400-2.png}
    % & \includegraphics[width=0.4\textwidth]{images/jfm-resp400-2.png} \\
    Re = 1800 $\omega=4.3$& \includegraphics[width=0.4\textwidth]{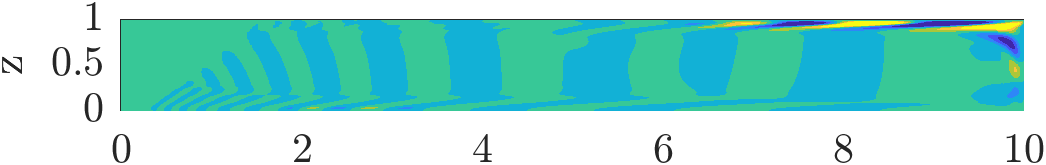}
    & \includegraphics[width=0.4\textwidth]{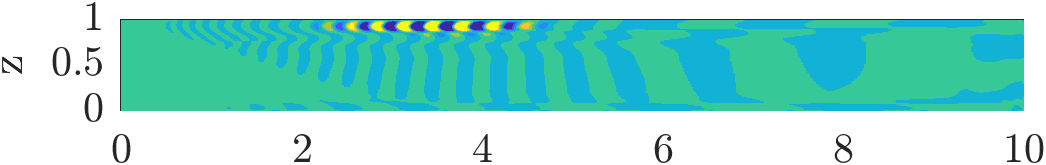} \\
    Re = 3000 $\omega=5.4$& \includegraphics[width=0.4\textwidth]{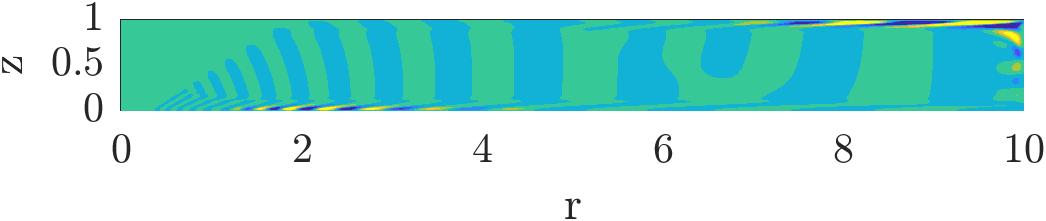}
    & \includegraphics[width=0.4\textwidth]{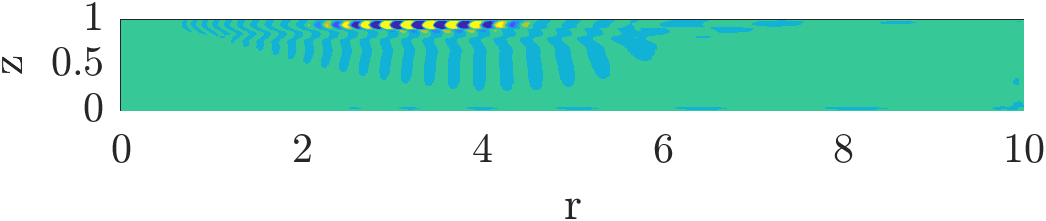} 
    \end{tabular}
 \caption{Azimuthal perturbation velocity $u_{\theta}(r,z)$ for the optimal forcing and response for $\omega\neq0$. For Re=250, 1800 and 3000 the optimal forcing is also optimal across all $\omega$'s. The colormap spans 8 equal subintervals of (-0.15 0.15) for the optimal forcing and (-0.6 0.6) for the optimal response. }% \end{table}
    \label{optforcingomesup1}
\end{figure}

The evolution of the optimally forced structure is now described as $Re$ increases beyond 250. It is first noted that, as shown in figure \ref{optgain} (right), the optimal gain $G_{opt}(Re)$ grows exponentially with $Re$. \review{Other examples of an exponential scaling of gain include a backward-facing step \citep{blackburn2008convective} and an oscillatory pipe flow \citep{xu2021non}. %fig 11b in blacknurn+barkley and fig 3 duo xu + avila 
} We note that the experimental study of 
% Gauthier {\it et al.}
\cite{Gauthier_jfm_1999}, performed for $\Gamma\approx20$, has suggested a supercritical bifurcation as the origin of the observed rolls. Our observations do not corroborate this hypothesis since the angular frequency content of the response features that a continuum of frequencies can be excited externally, at odds with the dominant frequency stemming from a Hopf bifurcation. Moreover, the steep exponential increase of $G(Re)$ reported above might be responsible, in experimental conditions where external forcing stays uncontrolled, for the apparent bifurcating behaviour where the amplitude of the response increases rapidly with $Re$. As $Re$ continues to increase, as shown in figure \ref{optforcingomesup1} the optimal forcing evolves with $Re$ from a wide support within the B\"odewadt layer to thinner structures in both the B\"odewadt and the shrouding wall, and even for $Re=3000$ also in the Ekman layer. For such high $Re$ values the respective supports of the forcing and the response are almost disjoint. \\
%This is again a common observation in non-normal flow where the mode and the its adjoint are localised further apart for growing non-normality.  (REF?)

%Secondly, as seen in figure \ref{optforcingomesup1}, the optimal forcing evolves with $Re$ from a wide support forcing in the B\"odewadt layer to thinner structures in B\"odewadt, but also in Ekman layer (e.g. $Re=3000$). The spatial support of response also changes. While for $Re\approx200$ forcing is located only a bit upstream form the response, for high $Re\approx3000$ supports of forcing and response are almost disjoint. This is again a common observation in nonnormal flow where the mode and the its adjoint are localised further apart for growing nonnormality.  
The results above unambiguously point towards an unsteady response in the shape of circular rolls for all $Re \gtrsim 150$. We emphasize that the crossing of the gain curves for $\omega=0$ and $\neq 0$ in figure \ref{optgain}(right) does \emph{not} define a threshold value for $Re$ because, as for any linear receptivity mechanism, the response depends linearly on the spectral content of the forcing history. Defining a threshold value for $Re$ is demanding because it is highly dependent on the amplitude levels that an experimentalist can detect in practice. By focusing on the value of the optimal gain ($G=4.9\times10^2$) at $Re=250$ the following
% hand-wavy %review 
evaluation can be made. Any parasitic vibration present in the experiment projected on the orthogonal basis of the optimal modes will have a nonzero component that will be optimally forced. If the amplitude of this component is, say, of order $O(10^{-2})$ it will be amplified by the linear mechanism by $\sqrt{G}\approx20$ to yield an $O(10^{-1})$ response, which can be detected in experiments. \\
% \lmw{I do not understand the 20 I would have found 7 for 0.1 and 2 for 0.01. If I am correct, i think it is more fair to go for 1 \% forcing and 2.2 response which remains O(1)}\\

%it is known that the forcing in the real experiment will never be in the shape of optimal forcing. To address this issue next section is devoted to suboptimal unsteady boundary forcing.  

% \textcolor{orange}{YD : PLAN : after $\omega=0$, we focus on $\omega \neq 0$
% In figure \ref{fig:4components}, we compute the optimal forcing and response
% for values of $\omega \neq 0$ and display all the three components as well as the pressure field. Th e fields have all been normalised such that their norm$\vert \vert \mathbf{\hat{u}} \vert \vert_{\mathbf{Q}}=\vert \vert \mathbf{\hat{f}} \vert \vert_{\mathbf{Q}}=1$. Strongest response always in the azimuthal velocity component. Analogy with planar case where streamwise streaks are the most amplified structures, there where streamwise=azimuthal.}\\

%-----then the non-zero omega frequencies

\subsection{Boundary forcing}

While the previous section offers an elegant formal manner to explain the circular rolls from (linear) optimality arguments, we note that at the experimentally relevant values of $Re$, the optimal forcing protocol corresponds to a force field that needs to be applied to the flow \emph{away} from the solid boundaries. Set-up imperfections are expected to induce forcing \emph{at} the boundaries rather than away from them. For this reason, it is unlikely that the observed linear response corresponds to truly optimal forcing, and it might be more relevant to concentrate on sub-optimal forcing.  Instead, we consider a forcing protocol which without being energetically optimal affects directly the flow through unsteady motion of the boundaries (and is hence not a suboptimal from the previous bulk-based optimisation).
% Such a protocol makes sense in the experimental context because of the imperfections of the driving motor that modify slightly the angular velocity of the rotor \lmw{I am worried that experimentalist are not going to like this point because the angular velocity is usually very well controlled. Should not we assume this forcing as the only one do able in practice in the code. This is similar to what Lopez and Serre did.}. Axisymmetric vibrations of the set-up are another \lmw{why the word another ?} possible cause of forcing. 
Modulations of the instantaneous (dimensional) angular velocity are considered in the form $\Omega(t)=\Omega_0(1+A\ \varepsilon(t))$, where $\varepsilon(t)$ represents a normalised unsteady forcing and $A \ge 0$ is a measure of its amplitude. This is similar to  \cite{Lopez_pof_2009}, \cite{Poncet_pof_2009} and \cite{do2010optimal}, except that $\varepsilon(t)$ is not monochromatic. The Reynolds number, now based on $\Omega_0$ rather than $\Omega$ (which depends on time), remains by definition unaffected by the value of $A$.
% We assume now that the (dimensional) rotation rate of the rotor is temporally as $\Omega(t)=\Omega_0(1+A\varepsilon(t))$, where $\varepsilon(t)$ represents a noise term with an amplitude $A \ge 0$.
% \agc{to be changed}

%In this section a forcing only applied at the domain boundary is considered. Such a forcing could be understood as present 

\subsubsection{Time integration}
The time modulation is simulated in practice by updating, at the end of each timestep after the prediction-projection step, the value of $v_{\theta}$ imposed on the shroud and the rotor as
\begin{equation}
\label{BCmodulation}
   % \Omega(t)=\Omega_0(1+A\varepsilon(t)),\\
    v_{\theta}(r,t)=r(1+A\ \varepsilon(t))
\end{equation} 
% \plq{seul le bandeau tourne... }
The modulation $\varepsilon(t)$ can be chosen in many ways. For most of this study, it is chosen as a Gaussian white noise of zero mean and standard deviation 1. $A$, which multiplies the white noise signal, is therefore the root mean square (rms) value of the forcing. All simulations are initiated with a zero perturbation field.

\begin{figure}
    \centering
    \includegraphics[width=0.8\textwidth]{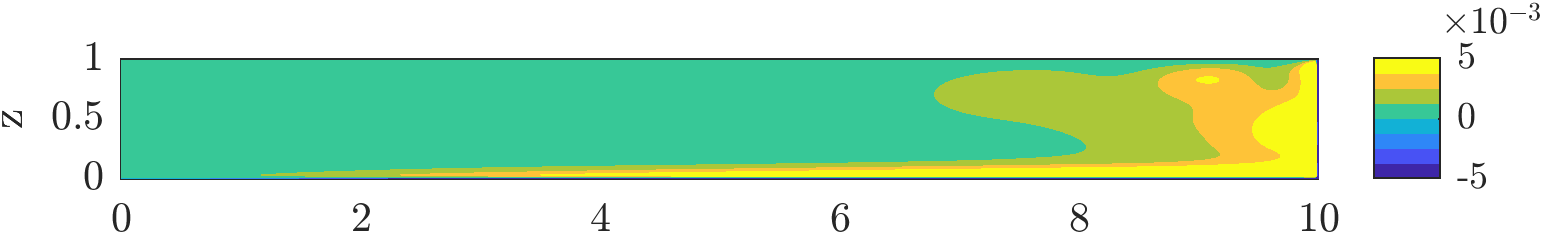}
    \includegraphics[width=0.8\textwidth]{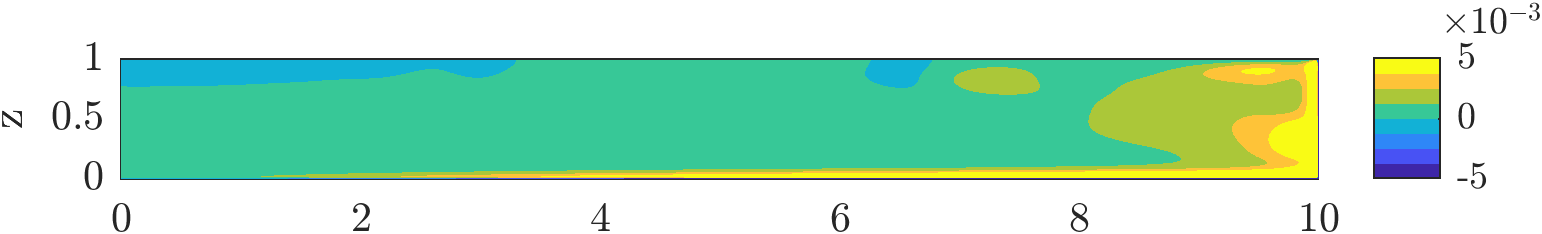}
    \includegraphics[width=0.8\textwidth]{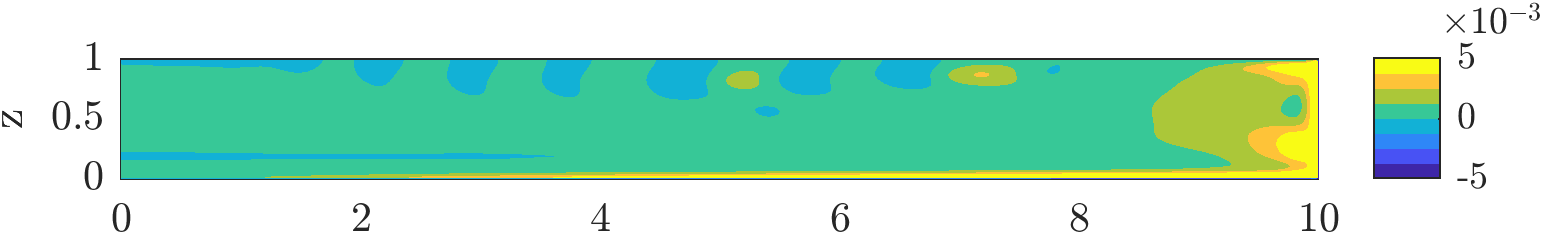}
    \includegraphics[width=0.8\textwidth]{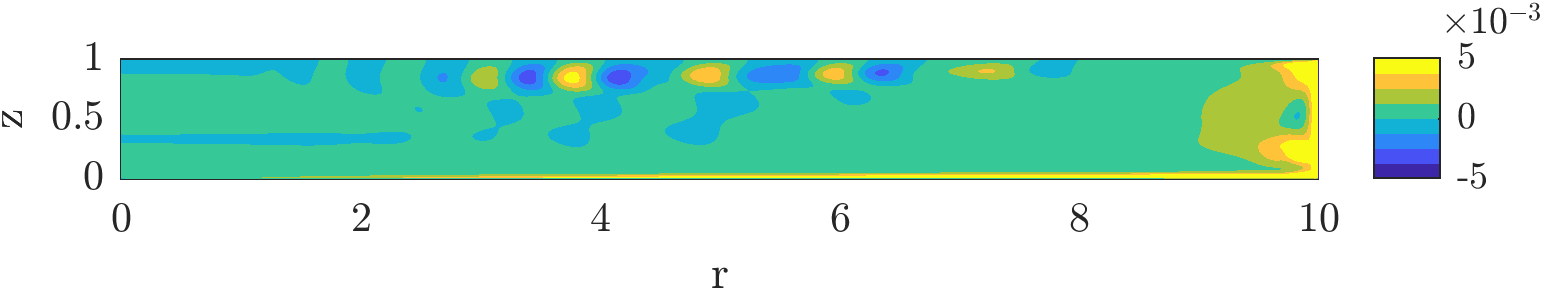}
    \caption{Nonlinear time integration in the presence of white noise forcing at the $z=0$ and $r=\Gamma$ boundary with amplitude $A=10^{-2}$. Snapshots of the azimuthal perturbation velocity $u_{\theta}(r,z)$. From top to bottom: $Re=70,\ 150,\ 250,\ 400$. A wavetrain of circular rolls is seen for $Re>250$. 
    % nl integration.% \textcolor{red}{YD : what is the value of $\omega$ here ?} this is white noise forcing from nonlin time integration. 
    % change this to snapshot from lin integration $\hat{u}_{\theta}$ response of the flow computed using the resolvent for $\omega=2$. Observe that the field values directly correspond to the values of $\hat{u}_{\theta}$ in the right panel of figure \ref{fig: ressol} for $\omega=2$.
    }
    \label{fig:nlresponsewhite}
\end{figure}

Figure \ref{fig:nlresponsewhite} shows representative azimuthal velocity snapshots obtained from nonlinear time integration of the forced flow. As will be clear later in section \ref{nonlin}, for the aspect ratio $\Gamma=10$ and $Re<300$ nonlinearity plays a negligible role and the observations are independent of whether the time integration is linear or nonlinear. This is why those results are discussed in the context of linearly optimal results. The response to the unsteady white noise forcing is localised, for low $Re\approx70$, next to the rotor and the shroud. For $Re \gtrsim 150$ and beyond, wavetrains of increasingly smaller circular rolls form inside the B\"odewadt layer, as is visible for $Re=250$ and $Re=400$. \\

%highly viscous flow  will be marked by low levels of gain 

Interestingly, the response to boundary forcing is similar to the response to the optimal forcing in the sense that the response consists also of wavetrains of circular rolls located inside the B\"odewadt layer. Due to the wide frequency content this response is however more widespread than in the case of optimal forcing (e.g. for $Re=250$ the wavetrain is detected for $r\in(2,7)$). A pronounced response near the rotor and especially the shroud is also seen, as a signature of the imposed forcing. 

%Similar character of the response, irrespective of the forcing being optimal or suboptimal, reinforced the analysis of the flow in the context of optimal linear analysis. 

\subsubsection{Resolvent approach to boundary forcing} \label{sec:resolventboundary}

The connection between the linearised time integration and the resolvent approach from \eqref{ioba} is the Fourier transform of the linearised governing equations. Selecting the additive forcing term $\mathbf{ \hat{f}}$ corresponding to the forcing protocol of \eqref{BCmodulation}, and then solving \eqref{ressol} for $\mathbf{\hat{u}}$ will therefore yield directly the Fourier transform of the response. It can be in turn compared against the Fourier transform of the probe signals extracted from linearised time integration. The same analogy has been exploited by \cite{cerqueira2014eigenvalue} in order to validate their input/output analysis. In our case, beyond immediate validation, this comparison is useful as it \review{shows the quantitative consistency 
% \sout{draws a link}
} between boundary forcing (although based on linear time integration rather than the nonlinear integration allowing for spectral mixing) and the inherently bulk-based resolvent approach.\\

\begin{figure}
    \centering
     \begin{minipage}{0.49\textwidth}
    \includegraphics[width=\textwidth]{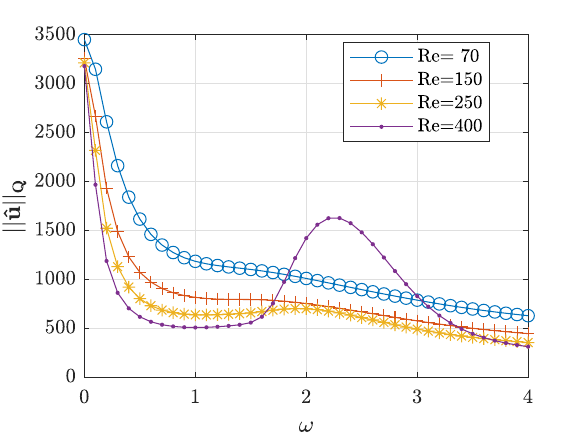}           
       \end{minipage}
       \begin{minipage}{0.49\textwidth}
           \includegraphics[width=1\textwidth]{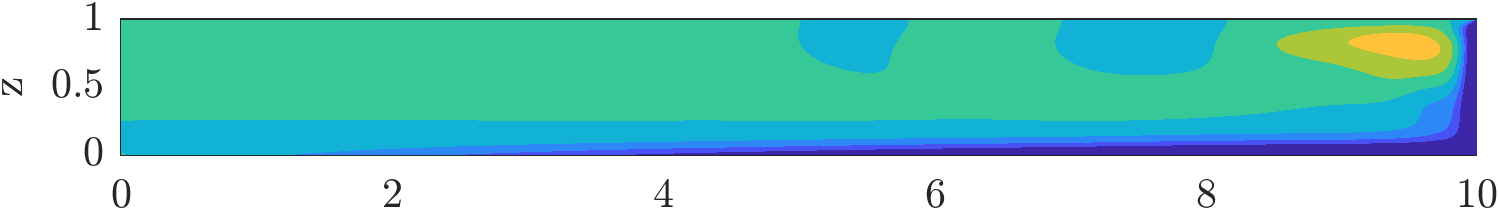} \newline \includegraphics[width=1\textwidth]{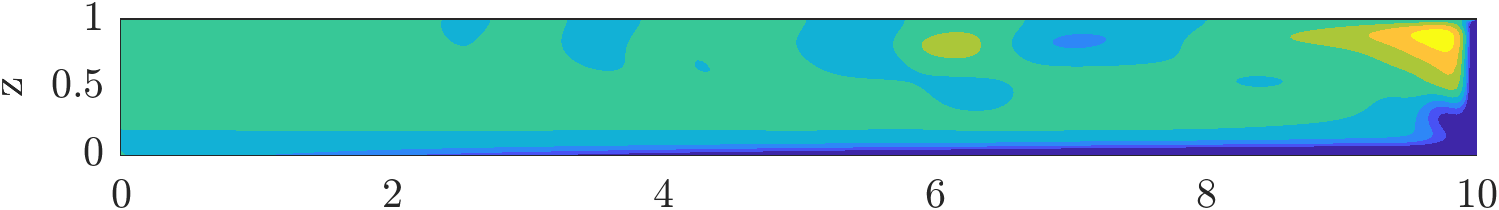}
    \includegraphics[width=1\textwidth]{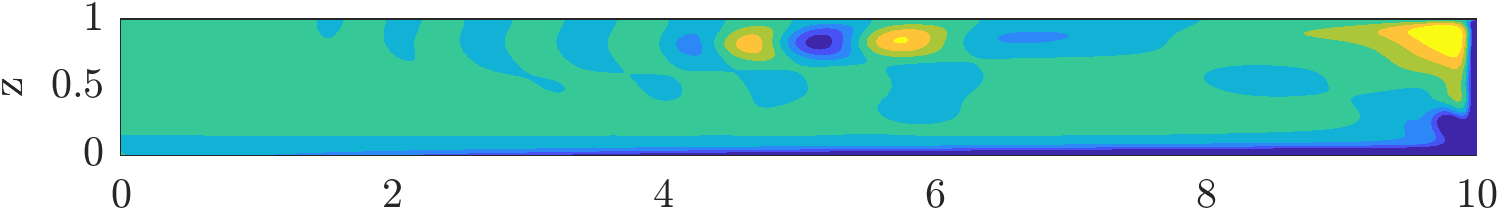}
    \includegraphics[width=1\textwidth]{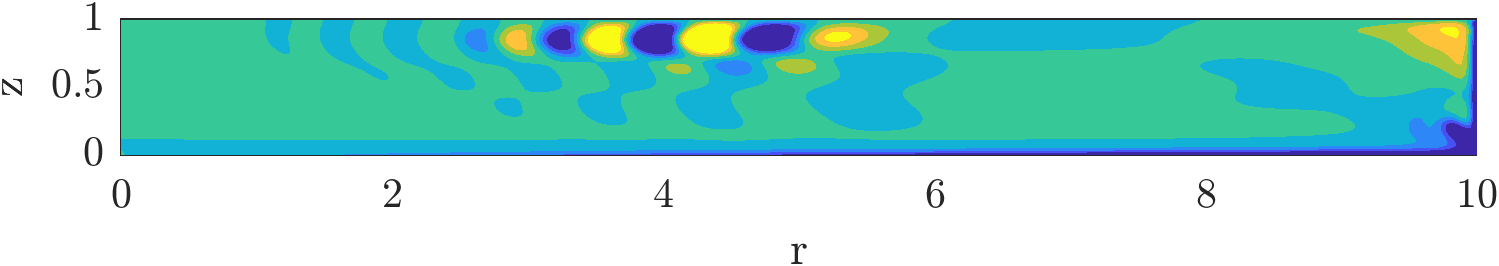}
       \end{minipage}
    \caption{Response to the boundary forcing obtained by the resolvent approach in section \ref{sec:resolventboundary}. Left: $\mathbf{Q}$-norm (squared) of the velocity response. 
    % compared with data obtained from optimal forcing. \lmw{Need to specify that there is two data plotted solid lines and symbols or the "compared" is misleading}
    Right, from top to bottom: real part of $\hat{u}_{\theta}(r,z)$ for $\omega=2$ for $Re=\ 70,\ 150,\ 250,\ 400$. The $\mathbf{\hat{u}}$ response is normalised so that $\vert \vert \mathbf{\hat{u}}\vert \vert _{\mathbf{Q}}=1$. Colormap spans $-0.3,(0.075),0.3$.}
    \label{fig8}
\end{figure}

% \ag{
We therefore select a specific, non-optimal 
% bulk-based
forcing term $\mathbf{P\hat{f}}$ corresponding to the Fourier transform of the boundary forcing \eqref{BCmodulation}. 
% This is achieved in practice by setting the azimuthal component 
% % $\hat{f}_{\theta}(r,z)=$
% to $r$ in the ghost cells 
% % at the outer edge of the domain,
% along the rotor and $\Gamma$ at the shroud.
% , for which $r \approx \Gamma$
% \lmw{not so easy to understand for me since it is said that resolvent applies only in the interior points}. 
Solving \eqref{ioba} with prescribed $\mathbf{P\hat{f}}$ yields $\mathbf{P\hat{u}}$, the real part of which is plotted in figure \ref{fig8}. The (squared) $\mathbf{Q}$-norm of $\mathbf{\hat{u}}$ is shown in the same figure. Due to the forcing being suboptimal, its dependence on $\omega$ differs from the gain curve in figure \ref{optgain} (left). 
% O%The optimal gain is proportional to $\mathbf{\hat{u}}^*\mathbf{Q}\mathbf{\hat{u}}$, which allows for direct comparison. 
% In particular for all values of $Re$ plotted in figure \ref{fig8}, the steady $\omega=0$ forcing yields the strongest response. 
Again, starting with $Re=250$ a hump visible in figure \ref{fig8}(left) around $\omega=2$ marks the preferred response of the flow in the shape of rolls. 
% \lmw{This paragraph is hard to understand }
\\ 
% }

%first fig 8

% The response to forcing with angular frequency $\omega$ \ags{to the white noise forcing} of the linear system, is computed using Eq. \eqref{ressol}, and is shown in Figure \ref{fig: ressol} both in the time domain (left) and in the frequency domain (right), for given values of $A$ and $Re$. 

% The response in the frequency domain can be interpreted as a gain curve comparable to figure \ref{optgain}.

% Figure \ref{fig:nlresponsewhite} shows arbitrary $u_{\theta}$ snapshots of the same nonlinear response for a given arbitrary forcing amplitude $A=10^{-2}$ and for increasing $Re=70,150,250$ and $400$. For the chosen value of $A$, stator patterns typical of forced rolls emerge for $Re=250$ and $400$.

\begin{figure}
    \centering
    \includegraphics[width=0.49\textwidth]{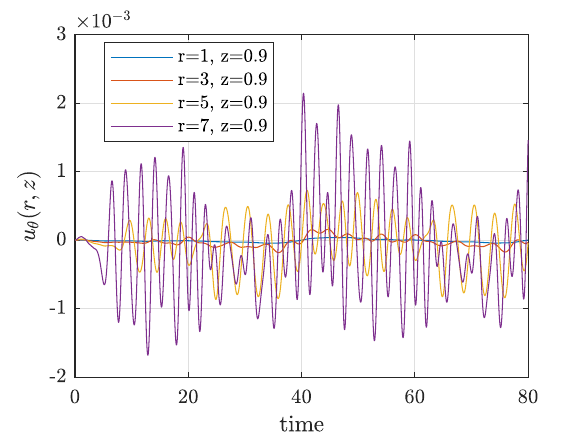}
    \includegraphics[width=0.49\textwidth]{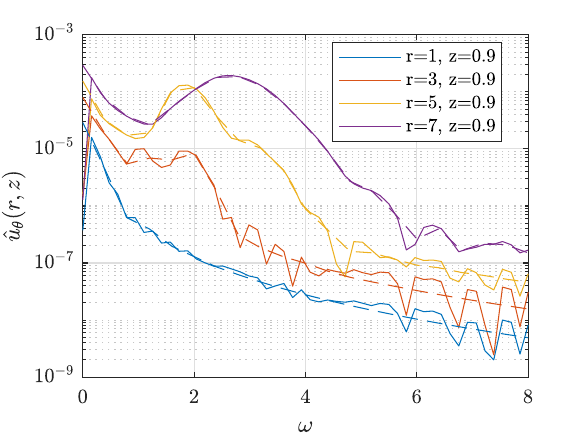}
    \caption{Response to boundary forcing by 
    Gaussian white noise with $A=10^{-2}$, Re=200.
    Left: time series of perturbation velocity $u_{\theta}$ obtained from time integration for different probes at varying $r$ inside the B\"odewadt layer z=0.9. Right: comparison between the time integration and the resolvent approach to boundary forcing in frequency space. Fourier amplitude spectrum of these time series (solid lines) {\emph versus} ${|\hat{u}_{\theta}|}$ associated with boundary forcing, computed from resolvent analysis (dashed lines) and evaluated at the same spatial positions.  Numerical parameters $dt=4\cdot 10^{-3}$, $nt=20\ 000$, $d\omega=0.157$, $\omega_{Nyquist}=785$ corresponding to the timestep, number of timesteps, sampling angular frequency and Nyquist angular frequency \review{(the highest angular frequency that can be reliably measured equals ${\pi}/{dt}$)} respectively. Only the later half of the signal was used for the calculation of the Fourier transform ($40<t<80$). }
    \label{fig: ressol}
\end{figure}

More insight into the preferred response frequencies of the flow is possible through the analysis of the probe signals extracted from linear time integration. The raw signals, corresponding to four radial positions $r=1,3,5,7$ along the B\"odewadt layer, are shown in figure \ref{fig: ressol}(left). The forcing is again white in time with equal amplitude for each frequency. The Fourier transform of the probe signal confirms, as already visible from the fields in figure \ref{fig:nlresponsewhite} and \ref{fig8}, that the strength of the response to boundary forcing depends on the radial position. The radially inwards flow in the B\"odewadt layer suggests that the distance to the shrouding wall, and therefore the varying thickness of the boundary layer, are the physically meaningful variables to explain this dependency, as also suggested by \cite{Gauthier_jfm_1999} (we will however stick to the variable $r$ for commodity). The preferred response frequency evolves also with the radial position. It is close to $\omega=0$ for the probe at $r=1$, but close to $\omega=3$ for the probe $r=7$. The Fourier amplitude can be directly compared to the pointwise amplitude of $\mathbf{\hat{u}}$, as mentioned earlier. The convincing overlap of both data (see figure \ref{fig: ressol}(right)) demonstrates the equivalence between the linear time integration and the direct resolvent solve based on \eqref{ressol}.

\review{An agreement between current results and the numerical study of \cite{do2010optimal} is noted when comparing figure \ref{fig: ressol} and figure 6 from \cite{do2010optimal}. Both spectra are characterised by a broad curve centered around $\omega=2.4$. The position of the rolls around the mid radius of the cavity also agrees while comparing figure 8 and figure 5 from \cite{do2010optimal}.   }

% \begin{figure}
%     \centering
%     \includegraphics[width=0.49\textwidth]{images/ps-200-25str.pdf}
%     \includegraphics[width=0.49\textwidth]{images/ps-200-121str.pdf}
%     \caption{PS Re=200 grid 300-80, norm of resol}
%     \label{fig:enter-label}
% \end{figure}
% \begin{figure}
%     \centering
%     \includegraphics[width=0.49\textwidth]{images/ps-1800-36str.pdf}
%     \includegraphics[width=0.49\textwidth]{images/ps-3000-36str.pdf}
%     \caption{PS Re=1800 \& 3000 grid 600-160, norm of resol, each point 600 sec}
%     \label{fig:enter-label}
% \end{figure}

%\subsection{Role of the pseudospectrum}

\subsection{Computation of the pseudospectrum}

\begin{figure}
    \centering
        \includegraphics[width=0.49\textwidth]{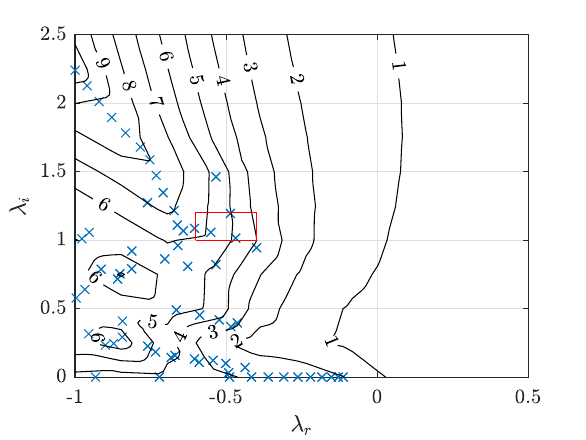}
        \includegraphics[width=0.49\textwidth]{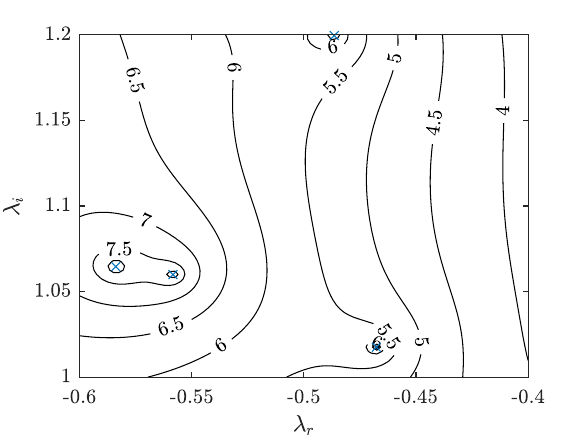}
        \includegraphics[width=0.49\textwidth]{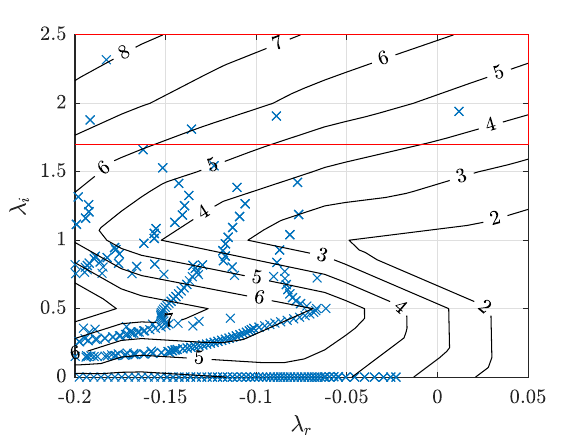}
        \includegraphics[width=0.49\textwidth]{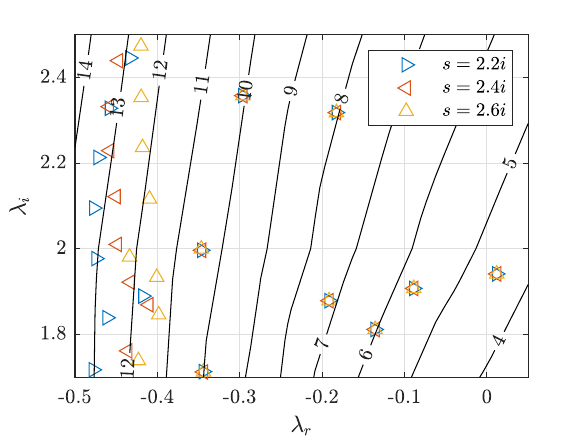}
    \caption{Pseudospectrum levels $log_{10}\vert \vert \mathbf{R} \vert\vert_\mathbf{Q}$ in the complex plane  where $\lambda_r$ and $\lambda_i$ are the real and imaginary part of the complex number $i\omega$. Top: $Re=200$, bottom: $Re=3000$. The right plot corresponds to the red inset in the left plot, where three different values of the shift $s$ in the shift-and-invert algorithm have been used. 
    % Left: $Re=200$, resolution R0. Right: $Re=3000$, resolution R1. Eigenvalues are marked with red symbols. Pseudospectrum is calculated in the points marked with black dots and linearly interpolated inbetween. 
    % \agc{sigma in legend}
    }
    \label{pseudo}
\end{figure}

A fundamental tool in the analysis of non-normal systems is the pseudospectrum, which is a generalisation of the concept of (eigen)spectrum. Recall that the spectrum of the linearised system \review{$\dot{{\bm q}}=\mathbf{L}{\bm q}$} is the (complex) set of eigenvalues of the associated linear operator. If a complex number $\omega$
belongs to 
the spectrum 
% $\sigma(\mathbf{R})$
$\sigma$, the resolvent $(i\omega\mathbf{I}-\mathbf{L})^{-1}$ (or expressed in the present case by \eqref{eq:res}) is undefined, 
% \lmw{\sout{by definition} ?}
 whereas it is continuous and smooth as a function of $\omega$ in the immediate neighbourhood of its pole at $i\omega$. Following \cite{trefethen2020spectra} the $\varepsilon$-pseudospectrum $\sigma_{\varepsilon}$ is defined as the complex set where the norm of the resolvent operator exceeds a given value $\varepsilon^{-1}$, with $\varepsilon$ a potentially small real number :
\begin{equation} \label{psdef}
     \sigma_{\varepsilon}=\{\omega \in \mathbb{C}, \vert \vert \mathbf{R} \vert \vert_{\mathbf{Q}}>\varepsilon^{-1}\}
\end{equation}
It includes the point spectrum and can be seen as its generalisation. In the current work, it is mainly used as an indicator of the strong non-normality of the underlying linearized operator. Owing to the definition \ref{psdef}, the cut of the pseudospectrum through the real axis directly yields the gain curves plotted in figure \ref{optgain}. The pseudospectrum is computed in each point, analogously to the optimal gains described in section \ref{optcomp}, except that $\omega$ is allowed to be complex. Note that such computations rely on the shift-and-invert algorithm itself dependent on a shift parameter $s$. Contours of the pseudospectrum are reported in figure \ref{pseudo}. As expected for a strongly non-normal operator, the isocontours of $\sigma_{\varepsilon}$ do not form concentric circles around the eigenvalues yet the contours encircle more than one eigenvalue. Still, very close locally to a given eigenvalue, the isocontours found for $Re=200$ form closed loops around specific eigenvalues, as can be seen in the top right panel in figure \ref{pseudo}. Upon increasing $Re$, the non-normality of the linearized operator increases and the background level in the pseudospectrum grows, as shown in figure \ref{pseudo}.\\

%---------Pseudospectrum figures-------

Another useful information from pseudospectra is the sensitivity of an eigenvalue to arbitrary perturbations of the operator $\mathbf{L}$ \citep{cerqueira2014eigenvalue,brynjell2017stability}. As shown in chapter 28 of \cite{trefethen2020spectra} for a convection-diffusion operator, high levels of non-normality typically cause iterative Arnoldi methods to converge to false eigenvalues. This is easily verified by comparing the effect of different shift values $s$ in the shift-inverse Arnoldi iteration. Here, as in \cite{cerqueira2014eigenvalue}, very large level of pseudospectral contours, typically above 11, prevent ARPACK from accurately converging all the eigenvalues. This algorithmic sensibility is not only a numerical convergence problem, it also highlights a physically ambiguous situation. Individual eigenvalues, when they are not robust, do not yield a specific contribution to the dynamics. In particular the associated eigenfrequencies are not resonant, and a large response to forcing is obtained even for forcing frequencies far from the eigenfrequencies \citep{trefethen1993hydrodynamic}. This is associated with the presence of a large hump (rather than narrow peaks) in the gain curve. In such a case, frequent in shear flows such as jets or boundary layers, one speaks of \emph{pseudo-resonance}, see e.g. \cite{garnaud2013preferred}. Non-robust eigenvalues are easily spotted as soon as their location in the complex plane depends on the shift $s$ chosen by the user, see figure \ref{pseudo}(bottom right). Conversely, the most unstable eigenvalue $\lambda_r=0.011,\ \lambda_i=1.95$ at $Re=3000$ in figure \ref{pseudo} can be trusted, and the growth rate $\lambda_r$ predicted by the computation of eigenvalues will correspond to the growth rate of instability upon time linearized integration. \\

%A few more observations can be made concerning pseudospectra. Because of the definition \ref{psdef}, the cut of the pseudospectrum on the imaginary axis gives the gain curves plotted in figure \ref{optgain}. This forms a conceptual link between pseudospectrum and the optimal gain in the flow.\\

Finally, the pseudospectra can also be post-processed to extract a lower bound on the maximal transient growth associated with given initial perturbations. Following Chap. 14 in \cite{trefethen2020spectra}, the Kreiss constant is the lower bound of the maximal growth of the initial perturbation and is defined by:
\begin{equation}
    \mathcal{K}=\sup_{\varepsilon>0}\frac{\alpha_{\varepsilon}}{\varepsilon}
\end{equation}
where $\alpha_{\varepsilon}$ is the pseudospectral abscissa (the maximal real part of the $\varepsilon$ contour in the complex plane). Computation of $\mathcal{K}$ for pseudospectra in figure \ref{pseudo} gives $\mathcal{K}_{Re=200}=1.4\times10^3$ and $\mathcal{K}_{Re=3000}=2.2\times10^8$, suggesting very large growth potential, as demonstrated by \cite{Daube_cf_2002} and \cite{gesla2023subcritical} for the same set-up.

% \begin{figure}
%     \centering
%     \includegraphics[width=0.49\textwidth]{images/z150-200-11-11.mat.pdf}
%     \includegraphics[width=0.49\textwidth]{images/z150-200-11-11.mat.zoom.pdf}
%     \includegraphics[width=0.49\textwidth]{images/z150-200-21-21.mat.pdf}
%     \includegraphics[width=0.49\textwidth]{images/z150-200-21-21.mat.zoom.pdf}
%     \includegraphics[width=0.49\textwidth]{images/z150-200-41-41.mat.pdf}
%     \includegraphics[width=0.49\textwidth]{images/z150-200-41-41.mat.zoom.pdf}
%     \includegraphics[width=0.49\textwidth]{images/z150-200-81-81.mat.pdf}
%     \includegraphics[width=0.49\textwidth]{images/z150-200-81-81.mat.zoom.pdf}
%     \caption{Pseudospectrum on few complex grids 11x11 ... 81x81 Resolution 150x40}
%     \label{fig:enter-label}
% \end{figure}

\section{Experimental comparison} \label{expcomp}

% +As shown in \cite{Faugaret_2020} forcing in disc harmonics.

In this section, an effort is made to reproduce the experimental conditions where axisymmetric rolls were reported. A detailed comparison with the literature results can be challenging because of the many different configurations of the rotor-stator system reported in the literature. Apart from the value of the aspect ratio $\Gamma$, special attention has to be given to whether or not the shroud is rotating and whether the cavity extends or not to the axis. The latter case corresponds to the presence of a hub. Relevant experimental studies of the rotor-stator configuration at moderate aspect ratio $\Gamma$ (from around 5 to around 20) are listed below :
\begin{enumerate}
    \item \cite{Gauthier_1998}, with a rotating shroud and no hub, reports circular rolls at $Re\approx70$ for $\Gamma=20.9$ and at $Re\approx180$ for $\Gamma=10$
    \item \cite{Schouveiler_jfm_2001} with a fixed shroud and no hub, reports circular rolls for a range of $\Gamma$, in particular at $Re\approx160$ (deduced from figure 3 \citep{Schouveiler_jfm_2001} for $\Gamma=10$)
    \item \cite{Poncet_pof_2009} with a fixed shroud and a rotating hub, reports circular rolls at $Re=160$ for $\Gamma=8.8$
\end{enumerate}
Numerical studies do not report sustained circular rolls in the absence of external forcing \review{\citep{Lopez_pof_2009, Poncet_pof_2009}}, at the exception of \citet{Daube_cf_2002} who considered larger $Re$ values.\\

We focus specifically on the article by \cite{Schouveiler_jfm_2001} and the corresponding parameters. The aspect ratio is hence temporary set to $\Gamma=8.75$, the shroud is fixed and the range of values of $Re$ $(0:300)$ is selected. Unsteady nonlinear simulations with boundary forcing are performed with three different forcing protocols. The main difficulties in comparing numerics to experiments are the fact that i) the amplitude of the forcing is unknown ii) the temporal frequency spectrum of the forcing is often unknown. Three qualitatively different types of forcing have hence been considered here~: monochromatic forcing, analogously to \cite{do2010optimal}, with specific forcing frequency $\omega=1$, harmonic forcing where only harmonics of $\omega=1$ are considered, and eventually white noise forcing. In the case of harmonic forcing all the temporally resolvable harmonics of the disc angular frequency are forced with equal amplitude. This type of forcing is especially relevant to the experimental comparison : as shown in both \cite{Gauthier_jfm_1999} and \cite{Faugaret_2020}, the main spurious perturbations present in the experimental set-up are the disc harmonics. 
%For simplicity, we fix to unity the amplitude of all harmonics of the rotating frequency. 
% The frequency of the response will however only include the leading harmonics because, as seen in figure \ref{fig: ressol}, any forcing with high angular frequency will be strongly damped.
The studies of the forced regimes focused mostly on  monochromatic forcing \citep{Gauthier_jfm_1999,Lopez_pof_2009,do2010optimal}. \\
% Whenever the response is linear those monochromatic forcing protocols can be seen as a subset of the white noise forcing.
\\

%with a radial downstream \ag{advection?} located instaneously

\begin{figure}
    \centering
    \includegraphics[width=0.77\textwidth]{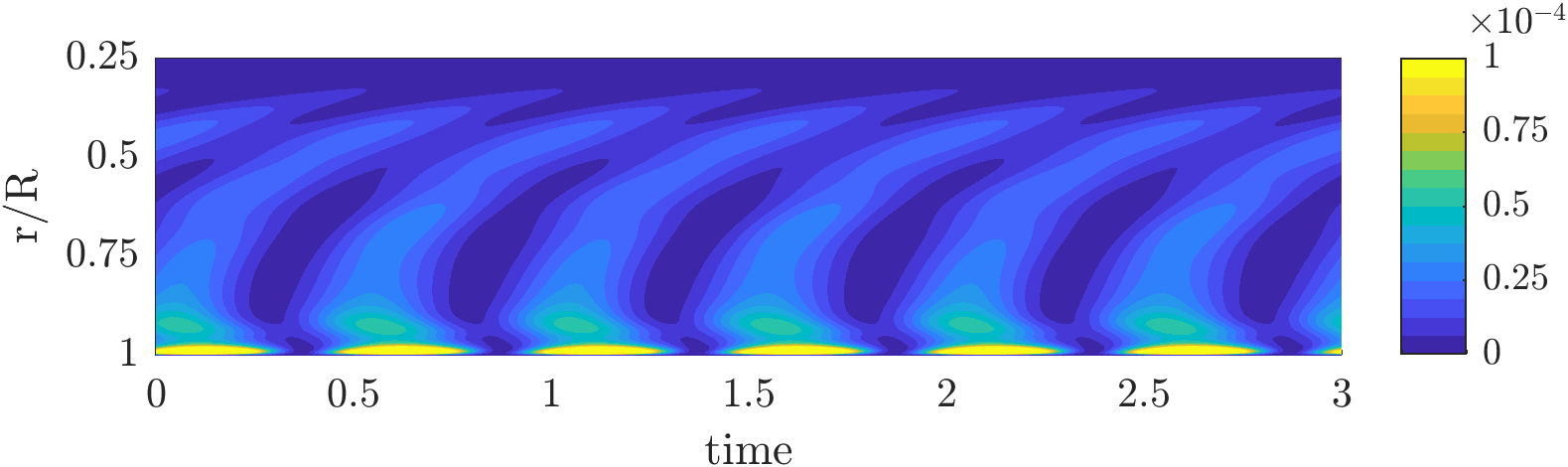}
    \includegraphics[width=0.22\textwidth]{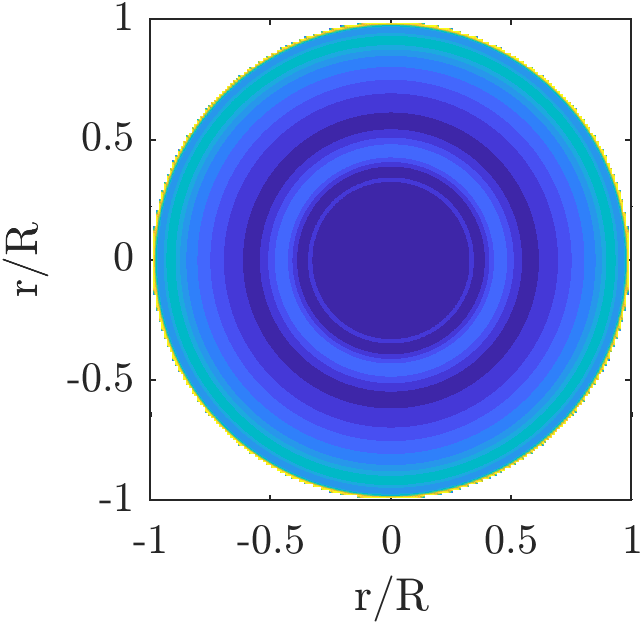}
%     \caption{Energy integral $E(r,t)$ at Re=225, $\Gamma=8.75$ for mono-frequency forcing at angular frequency $\omega=1$. Time is non-dimensionalised by the disc rotation period, so that the response has period unity. Excited rolls travel inwards with a radial phase speed depending on the radius. No merging and pairing of the rolls is observed. }
%     \label{fig:exp_mono}
% \end{figure}

% \begin{figure}
%     \centering
    \includegraphics[width=0.77\textwidth]{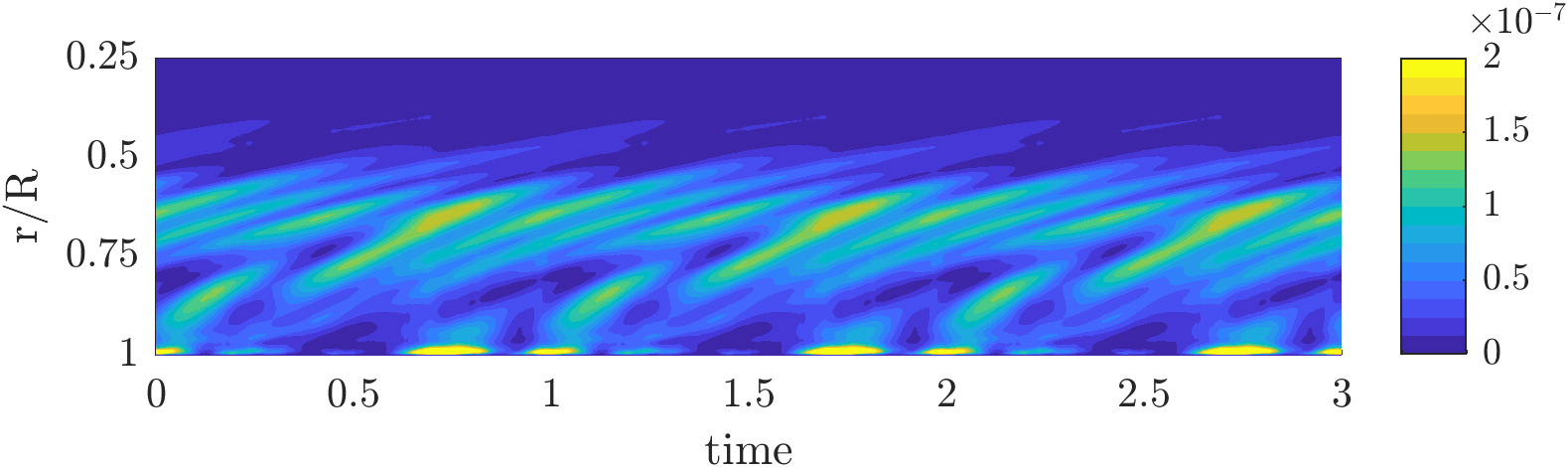}
    \includegraphics[width=0.22\textwidth]{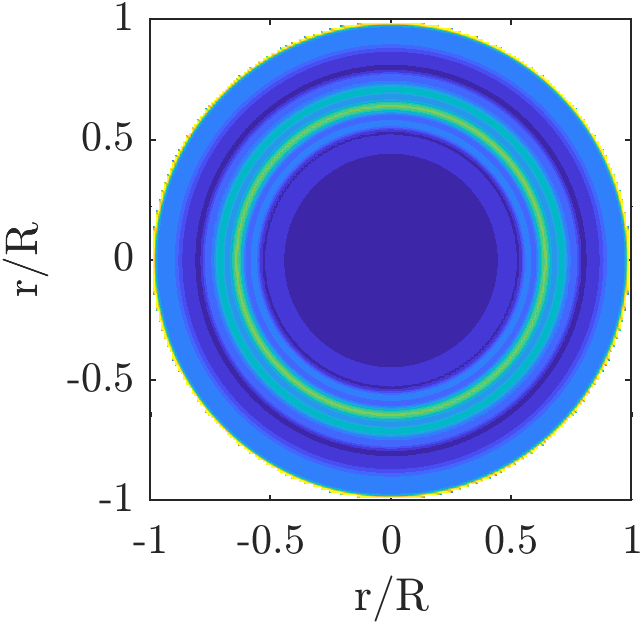}
%     \caption{Energy integral $E(r,t)$ at Re=225, $\Gamma=8.75$ for harmonic frequency $\omega=1,\ 2,\ 3, ...$ forcing. Time is non-dimensionalised \lmw{as in figure~\ref{fig:exp_mono} \sout{by the disc rotation period, so that the response has period unity}}. Pairing and merging of the rolls \lmw{are \sout{is}} observed here. Rolls merge because the phase speed depends on the radial position and the roll frequency $\omega$.
%     }
%     \label{fig:exp_harmo}
% \end{figure}

% \begin{figure}
%     \centering
    \includegraphics[width=0.77\textwidth]{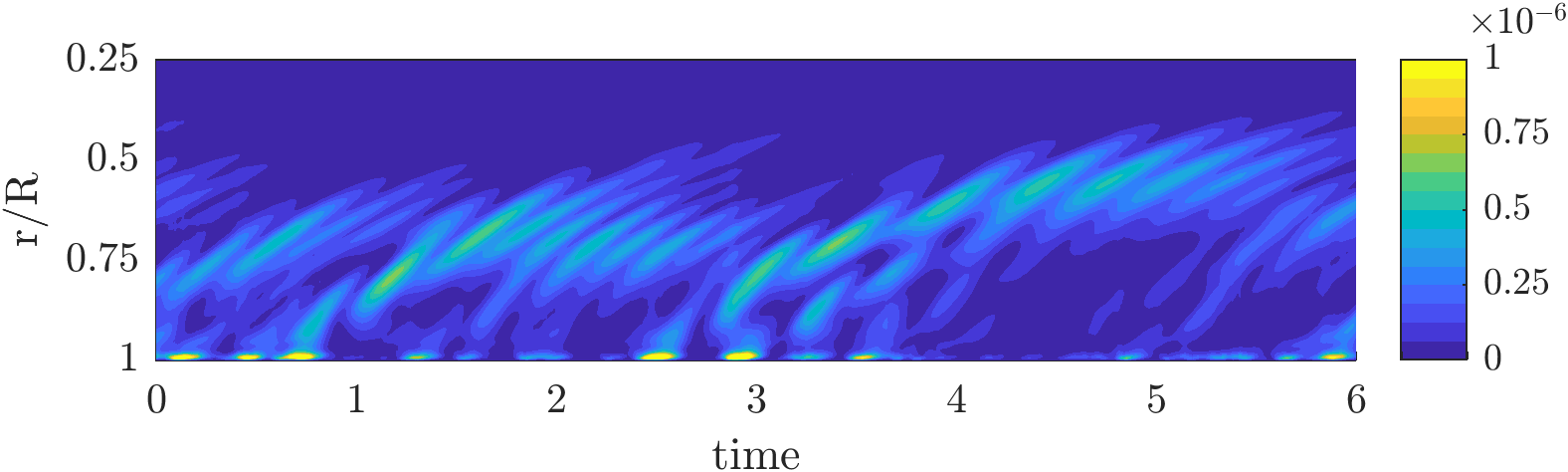}
        \includegraphics[width=0.22\textwidth]{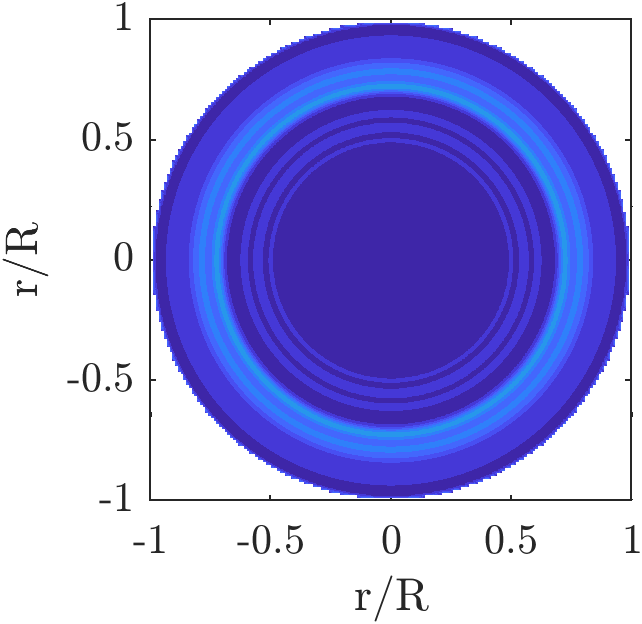}
    \caption{Energy integral $E(r,t)$ at Re=225, $\Gamma=8.75$. Time is non-dimensionalised by the disc rotation period. An initial transient was neglected. Excited rolls travel inwards with a radial phase speed depending on the radius. Top: mono-frequency forcing at angular frequency $\omega=1$.
     % , so that the response has period unity.
     No merging and pairing of the rolls is observed. 
     Middle: harmonic frequency $\omega=1,\ 2,\ 3, ...$ forcing. Pairing and merging of the rolls are observed here. Rolls merge because the phase speed depends on the radial position and the roll frequency $\omega$.
     Bottom: white noise forcing. Complex roll dynamics is observed due to all frequencies being excited by the forcing. }
    %\agc{Second space time to be deleted}}
    \label{fig:expForcing}
\end{figure}

For these three forcing protocols
an $r$-dependent observable needs to be monitored for all times. We select the observable $E(r,t)$ defined as 
\begin{equation}
    E(r,t)=\int_{\delta}^1((u_r)^2+(u_{\theta})^2+(u_z)^2) \ dz,
\end{equation}

where $\delta=0.2$ (in units of $H$) corresponds to integrating over 4/5 of the interdisc spacing (thereby excluding the parallel boundary layer along the rotor). This parameter $\delta=0.2$ is preferred over $\delta=0$ for the clarity of the resulting space-time diagrams. The results for the three type of forcing are presented in figure \ref{fig:expForcing}. All figures are accompanied by a snapshot of the same energy $E(r,t)$ at arbitrary time, seen from above in order to emphasize the visual comparison with experimental photographs. The three types of forcing all lead to wavetrains of circular rolls with $E$ locally stronger in the interval $0.34\le r \le 0.6$ \review{(radius non-dimensionalised with $R$ for consistency with \cite{Schouveiler_jfm_2001})}. The rolls reaching closest to the axis correspond to monochromatic forcing, while the front seems to remain further from the axis when the spectral content of the forcing is richer. %This suggests that nonlinear interactions in the central axial region conspire to suppress perturbations.\textcolor{red}{Is it sure or simply the signature of the vanishing velocity near the axis?} \ag{I would not recall nonlinearity here.}
The radial velocity of the rolls is always negative (oriented towards the axis) and can be estimated directly from the space-time diagrams. Even in the monochromatic case, their velocity is not constant and it depends on $r$ : since the B\"odewadt layer thickens as the perturbations migrate towards the axis where the radial velocity vanishes, the phase velocity of the rolls has to decrease, until the rolls can no longer sustain. The space-time diagram for the monochromatic case, qualitatively similar to Fig. 11{a} in \cite{do2010optimal}, illustrates this point particularly well. In the harmonic case, the coexistence of different frequencies leads to different radial velocities being possible at a given radius. The dynamics becomes more complex since rolls can now travel at different velocities, leading to consecutive rolls merging or splitting. The pattern shown in Fig. \ref{fig:expForcing} (middle), which shows periodic pairing and merging near $r \approx 0.75$ resembles in particular the experimental pattern shown in Figure 6 in Ref. \cite{Schouveiler_jfm_2001} at the same value of $Re$. \review{It also shares some similarity with the dynamics reported in Fig 11b of 
% Do {\it et al.} (2010)
\cite{do2010optimal}
where the forcing is only monochromatic but harmonics in the response are generated by nonlinearity.} As for the white noise forcing protocol, pairing and merging events appear more intermittently amidst otherwise periodic roll propagation.\\

The perfect agreement of the localisation of the rolls and the corresponding spacetime diagram with the experimental results of \cite{Schouveiler_jfm_2001} both reinforce the interpretation of circular rolls as the preferred response of the system to external forcing. 
%  \agc{schouv locking can be mentioned}

\section{Nonlinear receptivity} \label{nonlin}
Receptivity to external disturbances in fluid flows is traditionally investigated using linear
% toolbox of % review 
resolvent analysis, which is usually justified by the small perturbation velocities at play. This has the huge advantage that the responses to individual frequencies can be used as a basis to describe the global response to arbitrary forcing. Extensions of this formalism to finite-amplitude perturbations lead to new mathematical difficulties. Early efforts have focused on generalising the parabolised
% suitability %review 
\review{stability}
equations \citep{bertolotti1992linear,herbert1997parabolized}, which unfortunately is not an option for recirculating flows. A recent attempt to add nonlinear triadic interactions to the resolvent formalism was recently proposed by \cite{rigas2021nonlinear} for boundary layer flow. For the rotor-stator flow the importance of nonlinear interactions in the dynamics of the circular rolls was already pointed out by \cite{do2010optimal}, who measured a non-zero mean flow correction due to self-interaction of the unsteady rolls. 
% Similarly, the results from the space-time diagrams of the previous section suggest that nonlinear interactions might be at play in the merging/pairing phenomenon.
Moreover strictly nonlinear structures (i.e. without any linear counterpart) have been determined numerically for the unforced problem by \cite{gesla2023subcritical}. It is legitimate to enquire whether they can also be detected in the forced problem, at least in the presence of large enough forcing amplitudes. In the present section, we revisit the nonlinear time integrations from Section 4 for a larger range of the parameters $A$ (forcing amplitude) and $Re$.

\subsection{Effect of the nonlinearity}

\begin{figure}
    \centering
    \includegraphics[width=0.49\textwidth]{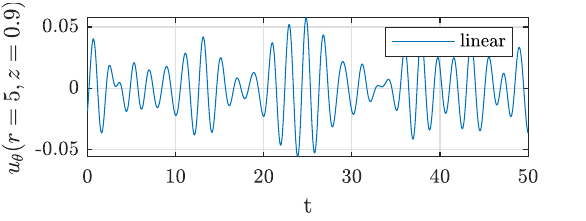}
    \includegraphics[width=0.49\textwidth]{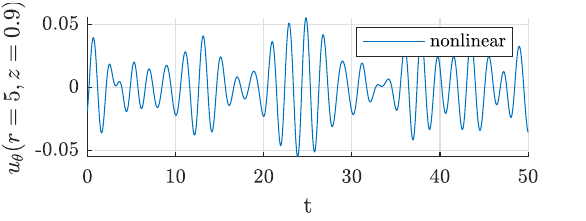}
    \includegraphics[width=0.99\textwidth]{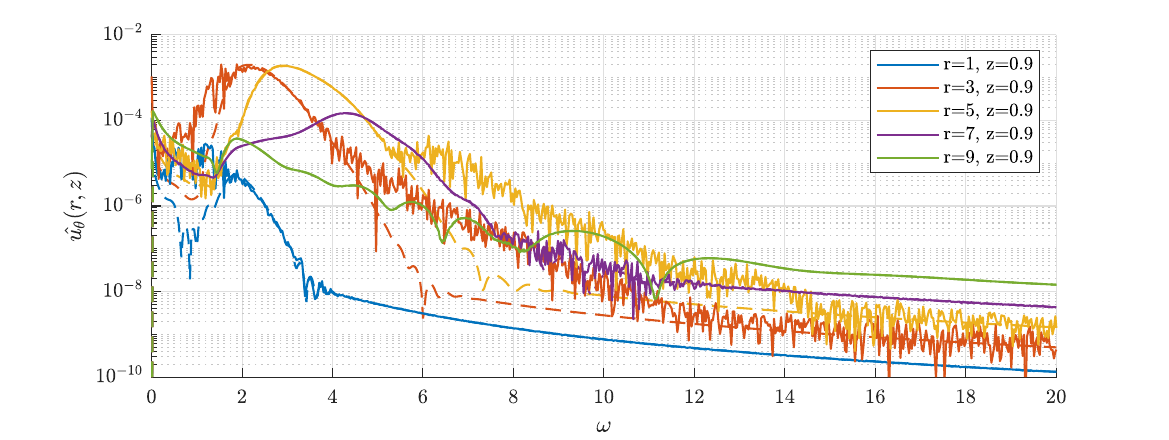}
    \caption{Comparison of linear and nonlinear time integration for $Re=600$, numerical resolution R1. Top: perturbation azimuthal velocity  $u_{\theta}$ at location (r=5, z=0.9) from linear (left) and nonlinear time integration (right). Bottom: Corresponding spectrum of non-linear (solid lines) and linear (dashed lines) velocity signal at  $z=0.9$ for various $r$.}
    \label{fig:compLNL600}
\end{figure}

\begin{figure}
    \centering
        \includegraphics[width=0.49\textwidth]{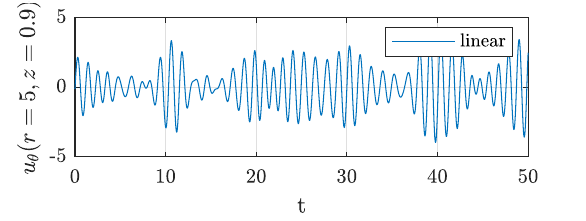}
            \includegraphics[width=0.49\textwidth]{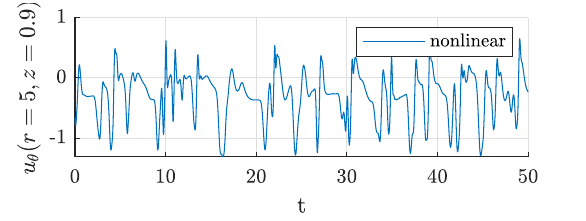}
    \includegraphics[width=0.99\textwidth]{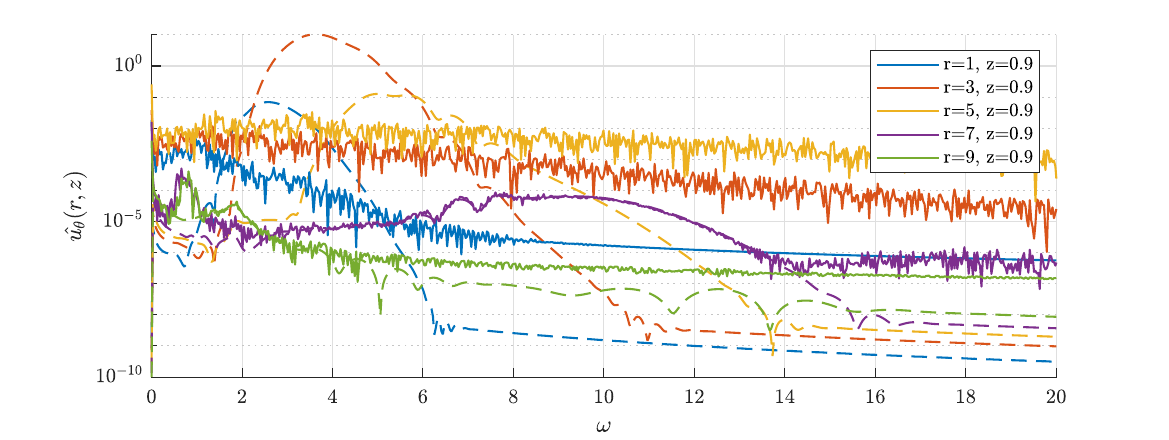}
    \caption{Comparison of linear and nonlinear time integration for $Re=2100$, numerical resolution R1. Top: perturbation azimuthal velocity $u_{\theta}$ at location (r=5, z=0.9) from linear (left) and  nonlinear time integration (right). Bottom: Corresponding spectrum of non-linear (solid lines) and linear (dashed lines) velocity signal at $z=0.9$ for various $r$.}
    \label{fig:compLNL2100}
\end{figure}

We begin by describing the effect of increasing the forcing amplitude on the gain curves such as fig. \ref{fig: ressol}.
Since nonlinear effects are expected to become more important with increasing $Re$, we first illustrate its effect by choosing two representative values of $Re=600$ and $2100$.  Rather than choosing two different values of the forcing amplitude in Eq. \ref{BCmodulation}, we choose only one such value $A=10^{-2}$ and compare respectively linear and nonlinear temporal simulations started from the same initial condition ${\bm u}(t=0)=0$. The case of $Re=600$ is investigated in figure \ref{fig:compLNL600} while $Re=2100$
 is analysed in figure \ref{fig:compLNL2100}. For $Re=600$, time series taken from probes located in the B\"odewadt layer reveal visually no difference between linear and nonlinear simulations. The Fourier transforms of these signals, interpretable as local gain curves, reveal however that the fastest scales (large $\omega \gtrsim 4$) are affected by nonlinear effects at least for the radii $r=3$ and $5$. At other radii neither the fast nor the slow scales are affected by nonlinearity. For $Re=2100$, the situation is clearer : the velocity time series look very different and their Fourier transforms, unsurprisingly, differ at all radial locations. In all cases nonlinearity manifests itself by a slower, exponential-like decrease of the Fourier amplitudes with increasing $\omega$, interpretable as a cascade in frequency space.
 % \lmw{not easy to get what is behind this idea}. 
 It is also nonlinear interactions between the frequencies that are responsible for the irregular shape of the spectrum.
 % sible for the presence of The Fourier distributions look noisy 
 This is because the timestep in the simulation cannot be commensurate with all frequencies present in the forcing.
 % \lmw{Looks a bit like we are not able to do the job correctly}
 \\
 

\begin{figure}
    \centering
                \includegraphics[width=0.5\textwidth]{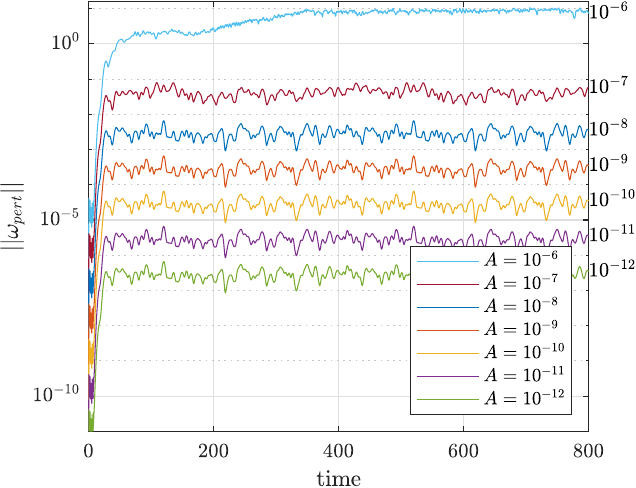}
                        \includegraphics[width=0.48\textwidth]{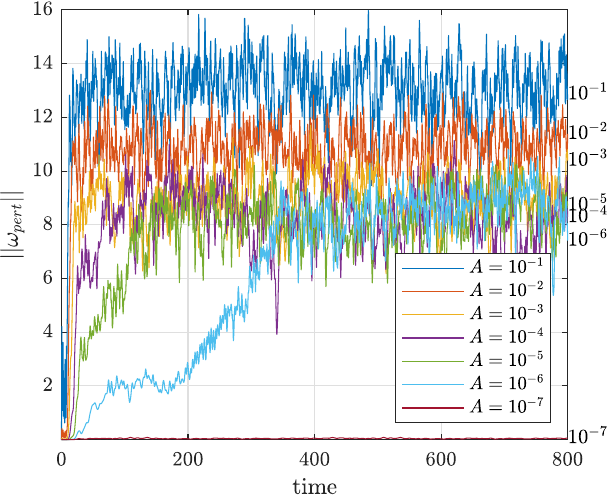}
    \caption{Global observable $\vert \vert \omega_{pert} \vert \vert$ in nonlinear time integration at $Re=2100$ for varying forcing amplitude $A$. The values of $A$ are indicated both in the legend and in the plot for simplicity. Starting from $A=10^{-7}$ nonlinear effects are observed. For $A\geq 10^{-6}$ the observable jumps to a different level corresponding to the presence of a non-trivial \emph{top branch} solution. }
    \label{fig:compLNLamp}
\end{figure}

Our goal is now to map in a $(A,Re)$ diagram the different behaviours encountered in the simulations, whether linear or nonlinear. The followed strategy is illustrated in figure \ref{fig:compLNLamp} by focusing on $Re=2100$. The time series corresponding to the observable $\vert \vert \omega_{pert} \vert \vert$ defined by 
\begin{equation}
    \vert \vert \omega_{pert} \vert \vert=\sqrt{\iint \vert\omega -\omega_{b}\vert^2 r dr dz},
\end{equation}
where $\omega$ is the azimuthal vorticity and $\omega_b$ that of the base flow, are shown in figure \ref{fig:compLNLamp} (left) in logarithmic scale, and in figure \ref{fig:compLNLamp} (right) in linear scale, for an amplitude $A$ spanning 6 decades from $A=10^{-12}$ to $A=10^{-6}$ (left) and 6 decades from $A=10^{-7}$ to $A=10^{-1}$ (right). For $A$ from $A=10^{-12}$ to $A=10^{-8}$, the time series in figure \ref{fig:compLNLamp} (left)
% turn out to be proportional to
differ only by
% are visually indiscernible except for 
their amplitude which turns out to be proportional
to the value of $A$
% \lmw{ Do we want to say that they would be visually indiscernible in the figure right so that we choose not to represent them ?}
. For larger $A=10^{-7}$ and especially $10^{-6}$, the time series change aspect and their amplitude no longer follows $A$ linearly. Note that the logarithmic scale is essential for this assessment, by comparison in figure \ref{fig:compLNLamp}(right) is much less easy to interpret along these lines.\\

\begin{figure}
    \centering
                \includegraphics[width=0.49\textwidth]{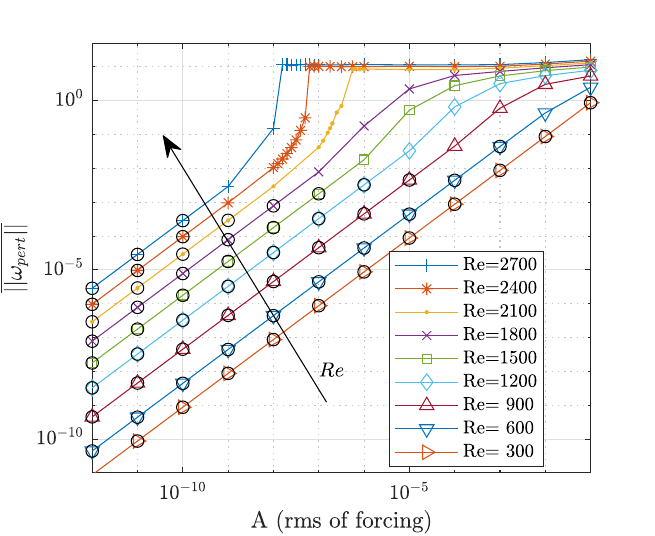}
                \includegraphics[width=0.49\textwidth]{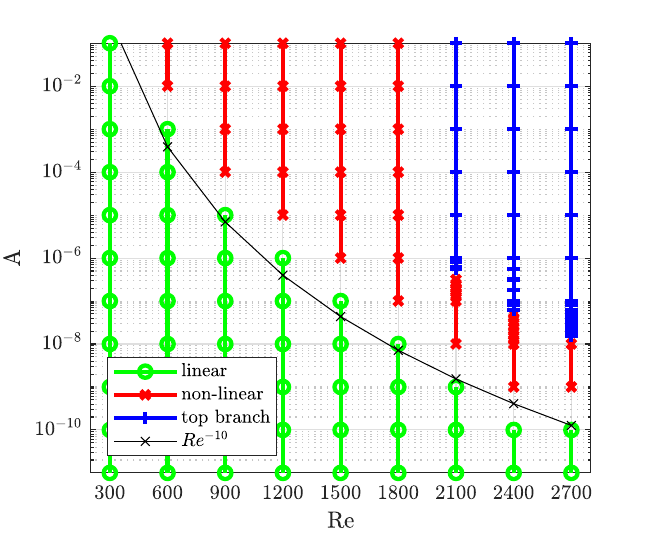}

    \caption{Left: mean observable value $\overline{\vert \vert \omega_{pert} \vert \vert}$ as a function of forcing amplitude $A$. The linear regime is indicated using  open symbols whenever the slope of the line is at most 1\% different from 1. For $Re\geq2100$ the mean observable value level jumps to the top branch level for strong enough forcing amplitudes. Right: amplitudes corresponding to linear and nonlinear regime indicated respectively with green and red symbols. The top branch is reached for $Re>1800$ whenever $\overline{\vert \vert \omega_{pert} \vert \vert}>5$.}
    \label{fig:compLNLAom}
\end{figure}

\begin{figure}
    \centering
\includegraphics[width=0.49\textwidth]{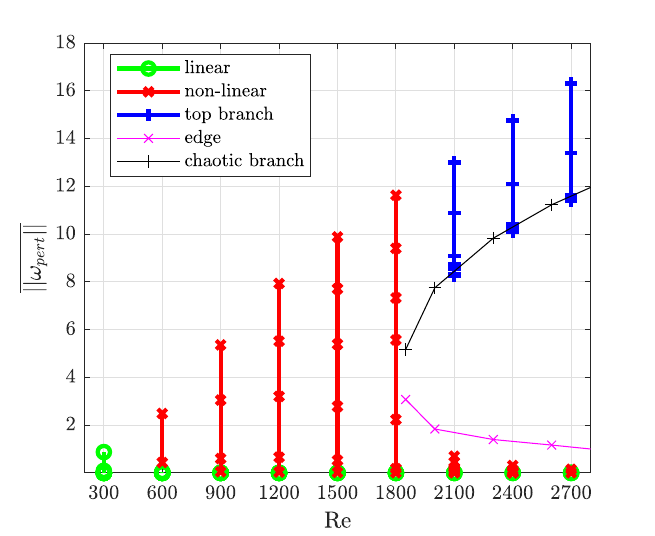}
\includegraphics[width=0.49\textwidth]{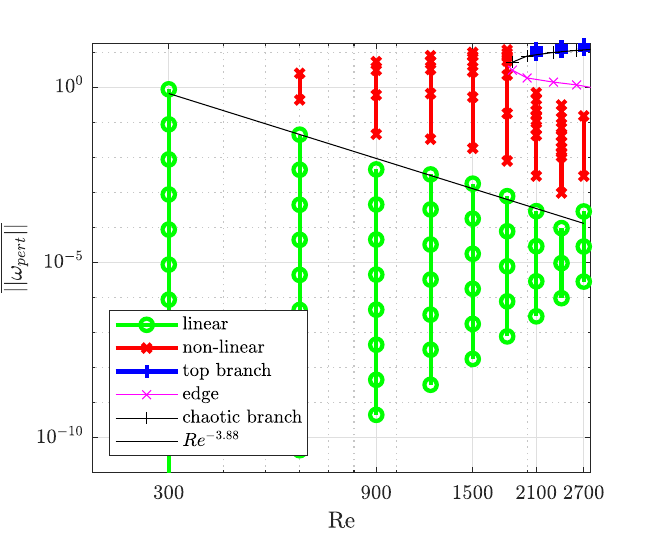}
                                                
    \caption{Left: mean observable value $\overline{\vert \vert \omega_{pert} \vert \vert}$ as a function of $Re$. The data corresponding to the chaotic solutions from \cite{gesla2023subcritical} are superimposed in respectively blue (top branch) and pink (for the edge branch). Same colour coding as in the previous figure. Right : same data plotted in logarithmic scales.}
    \label{fig:compLNLAom2}
\end{figure}

This analysis can be repeated for different values of $Re$. In figure \ref{fig:compLNLAom} (left), the time-averaged $\vert \vert \omega_{pert} \vert \vert$ is plotted directly versus the forcing amplitude $A$.
% with the time being implicit \lmw{do we need to specify that time is implicit ? Is there more than a mean ?}
This representation allows for a direct assessment of the parameter values for which $\vert \vert \omega_{pert} \vert \vert$ scales linearly with $A$, which suggests linear receptivity. All the points lying away from the related straight lines are labelled as nonlinear in the $(A,Re)$ map of figure \ref{fig:compLNLAom} (right). Beyond the simple linear/nonlinear labelling, we note that for the three largest values of $Re=2100,2400$ and $2700$ in figure \ref{fig:compLNLAom} (left), the observable $\vert \vert \omega_{pert} \vert \vert$ saturates with increasing $A$ to an almost constant level. This suggests that the final state of the simulation is the same whatever $A$, in other words that the forced system is attracted towards a different region of the state space for at least $Re \gtrsim 2100$. A non-trivial chaotic attractor was reported, for the same parameters, for the unforced problem, by \cite{Daube_cf_2002} and \cite{gesla2023subcritical}, for $Re \gtrsim 1800$. This attractor is different from the laminar one and it has a different attraction basin in terms of initial condition. There is little doubt that the same attracting chaotic state is identified here in the presence of strong amplitude forcing (the initial condition being fixed). A closer zoom on the figure, especially in linear scale, would show that the saturated value of the observable depends weakly on $A$, which suggests that the associated attractor is gently sensitive to the unsteady forcing. When such an attractor is reached, the corresponding point $(A,Re)$ in the map of figure \ref{fig:compLNLAom} (right) changes from red to blue. The points that are red correspond to the values of $(A,Re)$ where nonlinear effects do bend the gain curves (left figure), but no finite amplitude state is present and thus no saturation to any non-trivial state can occur. This leads to a 3-colour cartography of the parameter space $(A,Re)$ for the chosen forcing protocol parameterised by the amplitude parameter $A$. The notion of double threshold, popular in the shear flow transition community \citep{grossmann2000onset}, appears here in both $A$ and $Re$ (rather than in initial perturbation amplitude and $Re$), for a forced rather than unforced problem. The boundary between linear and nonlinear behaviour in \ref{fig:compLNLAom} (right) is consistent with an approximate fit of the form $A=O(Re^{-\alpha})$, with $\alpha \approx 10$. This large value of $\alpha$ is the signature of a very steep boundary. As for the boundary between red (nonlinear) and blue (non-trivial saturation), the present data does not suggest any simple fit. Besides the dataset is known to end at $Re \approx 3000$, beyond which all points are blue because the base flow is linearly unstable and the turbulent state is the only attractor left \citep{gesla2023subcritical}. Rather than a cartography in the $(A,Re)$ plane, where $A$ is specific to a given forcing protocol, we plot in Figure 
\ref{fig:compLNLAom2} the time-averaged observable $\overline{\vert \vert \omega_{pert} \vert \vert}$ versus $Re$, the value of $A$ being treated implicitly. The data is represented using the same colour coding as the preceding figure, both in
% \lmw{\sout{lin-lin (left figure) and in log-log coordinates (right figure)}
linear (left figure) and in logarithmic scales (right figure). In both plots the data corresponding to the non-trivial top and lower branches (i.e. edge states) obtained in \cite{gesla2023subcritical} are also superimposed for clarity (the same observable was used, see their figure 19). The two upper and lower branches from the unforced problem were reported to merge in a saddle-node bifurcation at $Re=Re_{SN} \approx 1800$. Interestingly, none of the red or blue data points, which correspond to the unforced problem, fall in the gap between the lower and upper branch. 
\review{This suggests that the data from the deterministic problem can be used to delineate between the two types of response, even in the presence of a finite-amplitude forcing. }
%To our knowledge it is the first time that deterministic data from an unforced problem compares so favourably with data from the forced problem. \yd{In particular we are not aware of any forced nonlinear fluid problem where there is evidence for the role played by non-trivial solutions of the unforced problem.} 
The right panel of the figure shows the same data in logarithmic scale. There, for the choice of this observable, the boundary between linear and nonlinear behaviour obeys an approximate power-law scaling $\overline{\vert \vert \omega_{pert} \vert \vert}=O(Re^{-\beta})$, with $\beta \approx 3.88$.\\

\subsection{Leaky chaotic attractors and finite lifetime dynamics}

\begin{figure}
    \centering
      \includegraphics[width=0.49\textwidth]{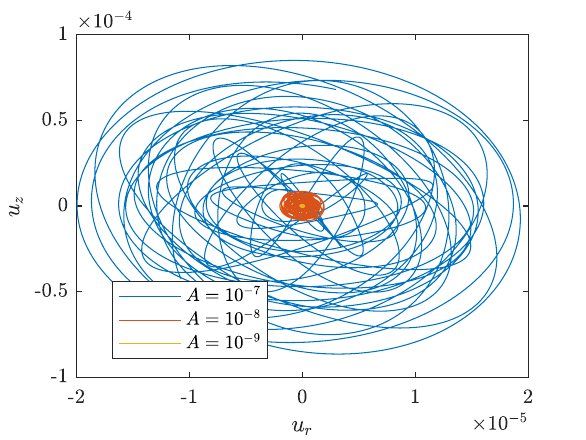}
      \includegraphics[width=0.49\textwidth]{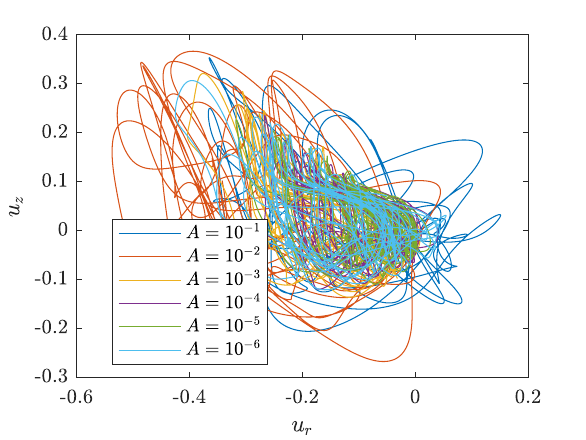}
    \caption{Two-dimensional state portrait $(u_r,u_z)$ from a velocity probe located $(r=5,\ z=0.9)$ at $Re=2100$. The width of the state portraits increases linearly with the forcing amplitude for $A\leq 10^{-7}$. For $A\geq10^{-6}$ state portraits lose their elliptic shape. The center of mass of the state portrait also shifts away from 0, indicating strong nonlinear interaction. }
    \label{fig:probe2100}
\end{figure}

From the previous subsection, we know that forcing the nonlinear problem with a sufficientlty large amplitude leads, for $Re$ above $Re_{SN} \approx 1800$, to a response different from the response triggered at lower forcing amplitudes. The difference is attested for global observables such as in figure \ref{fig:compLNLamp} and it is also clear from local velocity probes, as shown in Figure \ref{fig:probe2100} for $Re=2100>Re_{SN}$. In the left panel, low amplitude values $A \le 10^{-7}$ (corresponding to green points in Figure \ref{fig:compLNLAom}) lead to a disordered response whose diameter increases (linearly) with the value of $A$. In the right panel, only values of $A \ge 10^{-6}$ are displayed and it is apparent again, especially from the numbers along the axes, that the response depends weakly on $A$. This last point suggests the existence of an attracting state independent of $A$, in other words a deterministic attractor, solution of the forced problem, identical to the top-branch solutions identified by \cite{gesla2023subcritical}. Similar conclusions can be drawn for all $Re$ larger than 2100, including the supercritical values larger than $Re_c\approx 3000$ when only the non-trivial attractor is stable. \\

\begin{figure}
    \centering
    \includegraphics[width=\textwidth]{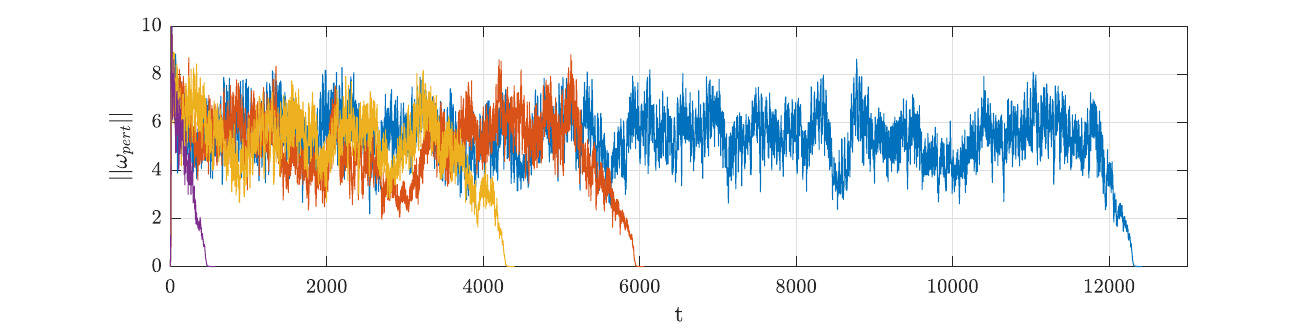}
    \caption{Time series of  $\vert \vert \omega_{pert} \vert \vert(t)$ for $Re=1840$, unforced problem. Random initial perturbations of amplitude $8\times10^{-3}$ added to the $u_{\theta}$ component. Four different initial conditions lead to four different lifetime values. }
    \label{fig:metastable}
\end{figure}

\begin{figure}
    \centering
                \includegraphics[width=0.79\textwidth]{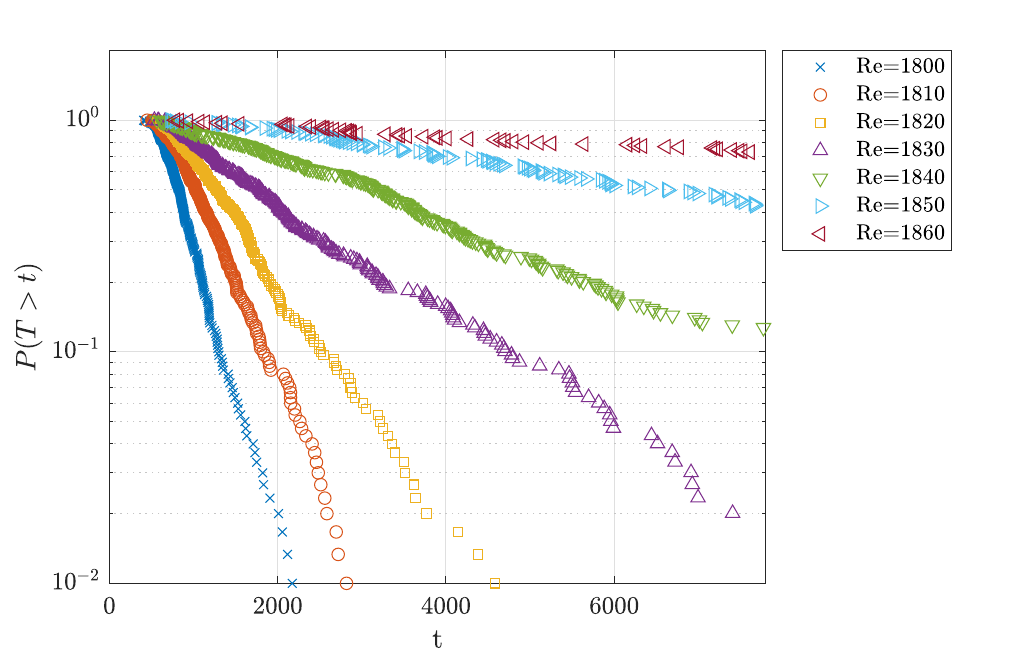}
    \caption{
    % Left: 
    Cumulative distribution function $P(T>t)$ for the lifetime $T$
    % \sout{probability $P$ for the lifetime $T$ to be larger than a given time}
    . $P$ is estimated as the fraction of runs that did not relaminarise in the time interval $(0:t)$. Relaminarisation is decided whenever $\vert \vert \omega_{pert} \vert \vert<10^{-3}$. For each value of $Re$ simulations with different random initial perturbation were conducted. 
    % Right: Dependence of the mean lifetime $\overline{T}$ on $Re$. The different data illustrate the sensitivity of the mean lifetime w.r.t. 
    % the total number of runs.
    % For a memoryless process the distributions in the left plot would be exactly exponential and the mean lifetime would be the reciprocal of their slope. 
    % \sout{Using three datasets of different size give an qualitative information about mean lifetime evaluation sensitivity.}
    }  
    \label{fig:lifetimestats}
\end{figure}

Departures from this picture can however be noted at lower $Re$ close (but above) $Re_{SN}$. Figure \ref{fig:metastable} shows the outcome several temporal simulations of the unforced problem for $Re=1840$, focusing on the time series of $\vert \vert \omega_{pert} \vert \vert (t)$. These time series demonstrate that the chaotic behaviour lasts for a finite time only, after which the observable rapidly reaches zero, indicating a return to the base flow. Some of the measured lifetimes are much longer than the duration of the return to the base flow, which justifies the label \emph{supertransient} \citep{lai2011transient}. The statistics of the lifetimes have been gathered over many such realisations of the deterministic problem, all initialised by different velocity fields drawn randomly.
The cumulative distribution of the lifetimes $P(T>t)$, where $T$ is the lifetime, are shown in figure \ref{fig:lifetimestats}
% (left)
 for several values of $Re$ between 1800 and 
%  1840
 \review{1860}. The distributions all look exponential for up to two decades, which suggests a memoryless process \citep{bottin1998statistical}.\\
 
%  The average value $\overline{T}$ of the lifetime is then plotted versus $Re$ in figure \ref{fig:lifetimestats}(right), where the $y$-axis is represented in logarithmic scale. The mean lifetime $\overline{T}$ increases fast with $Re$. Although no functional fit emerges from the data, the growth of $\overline{T}(Re)$ appears faster than exponential. %review

\begin{figure}
    \centering
    \includegraphics[width=0.49\textwidth]{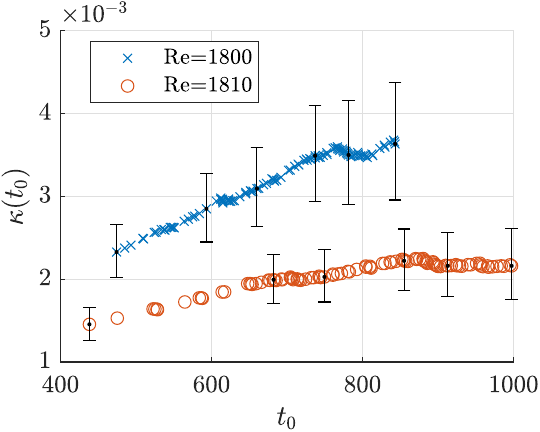}
    \includegraphics[width=0.49\textwidth]{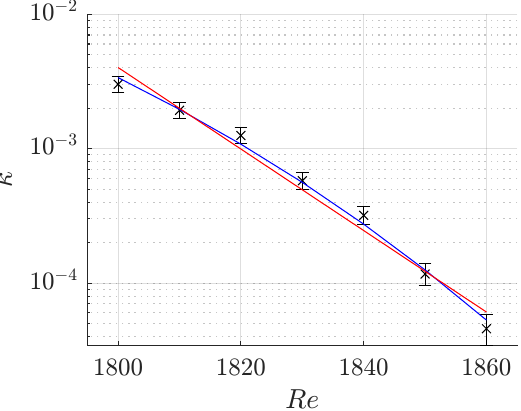}
    \caption{Left: Escape rate $\kappa(t_0)$ for $Re=1800$ and $Re=1810$. 
    The exponential distribution is reached when $\kappa$ is independent of $t_0$, which can be assumed for $t_0>800$ when$Re=1800$, and for $t_0>850$ when $Re=1810$. Right: dependence of $\kappa$ on $Re$. 
    % Due to an approximately exponential dependence the mean lifetime will depend exponentially on $Re$. 
    The fits in the figure are $\exp(-6.97\times 10^{-2}Re+   1.2\times 10^{2})$ (red) and  $\exp(-\exp(9.13\times 10^{-3}Re  -1.5\times 10^{1}))$ (blue), suggesting superexponential dependence on $Re$. }
    \label{fig:lt-plot-rev}
\end{figure}

\review{As noted by \cite{avila2010transient}, the perturbations need a certain time $t_0$, before they reach a leaky attractor. The mean lifetime $\tau$ of only these perturbations that have reached the leaky attractor can be computed by discarding all of the simulations that relaminarised before the time $t_0$: 
\begin{equation}
    \tau(t_0)=\left<t-t_0 | t>t_0\right>
\end{equation}
where $\left<\cdot\right>$ stands for the mean. $\tau(t_0)$ that is a constant function of $t_0$ is a characteristic feature of a memoryless process. \\
In case of relaminarisation of every simulation out of $r$ total simulations, the Maximum Likelihood Estimator (MLE) of $\tau$ can be computed using
\begin{equation}
    \tau=\frac{1}{r}\sum_{i=1}^rt_i.
\end{equation}
where $t_i$ is the relaminarisation time of individual simulations. 
However, because of the finite numerical integration time, not all simulations relaminarise. To account for the perturbations whose lifetime exceeds a censoring time $t_r$, the MLE is modified as \citep{lawless2011statistical}
\begin{equation} \label{mle}
    \tau=\frac{1}{r}\left( \sum_{i=1}^rt_i+(n-r)t_r\right),
\end{equation}
with $r$ thenumber of simulations that relaminarised and $n$ the total number of simulations. A 95\% confidence interval of the estimator \eqref{mle} is
\begin{equation} \label{mle-conf}
    \tau \times \left[ \frac{2r}{\chi^2_{2r,0.975}},\frac{2r}{\chi^2_{2r,0.025}}\right].
\end{equation}
where $\chi^2_{m,p}$ is the $p$th quantile of the chi-squared distribution with $m$ degrees of freedom.\\
The total number of simulations performed for each $Re$, together with $t_0$, the escape rate from the saddle $\kappa$ \citep{tel2008chaotic}, its confidence interval and a corresponding $\tau$, are all reported in table \ref{tab:ltt0}. The dependence of $\kappa$ on $t_0$ and $Re$ is shown in figure \ref{fig:lt-plot-rev}. Linear dependence $log(\kappa(Re))$ would suggest an exponential scaling of the mean lifetime as a function of $Re$, at least over the interval in $Re$ where the data was gathered (which lies well below the linear instability threshold).\\
\begin{table}
    \centering
    \begin{tabular}{ccccccc}
         $Re$ & $n$ & $r$ & $t_0$ & $\kappa(t_0)$ & $\kappa$ 95\% conf. int.& $\tau$ \\
%         1800&299&299&721.3&3.016e-03&[2.611e-03,3.449e-03]&331.6\\ 
% 1810&299&299&872.7&1.929e-03&[1.670e-03,2.206e-03]&518.4\\ 
% 1820&276&276&1057.7&1.240e-03&[1.064e-03,1.430e-03]&806.1\\ 
% 1830&183&181&1857.9&6.884e-04&[5.467e-04,8.461e-04]&1452.7\\ 
% 1840&298&261&2159.5&3.193e-04&[2.719e-04,3.704e-04]&3132.1\\ 
% 1850&58&37&741.2&1.315e-04&[9.259e-05,1.771e-04]&7604.1\\ 
% 1860&29&11&1072.6&6.970e-05&[3.479e-05,1.165e-04]&14347.0\\ 
1800&299&299&721.3&3.016e-03&[2.611e-03,3.449e-03]&331.6\\ 
1810&299&299&872.7&1.929e-03&[1.670e-03,2.206e-03]&518.4\\ 
1820&299&299&1026.1&1.253e-03&[1.085e-03,1.433e-03]&798.3\\ 
1830&298&293&1286.5&5.775e-04&[4.989e-04,6.618e-04]&1731.6\\ 
1840&298&261&2159.5&3.193e-04&[2.719e-04,3.704e-04]&3132.1\\ 
1850&198&114&722.6&1.169e-04&[9.640e-05,1.393e-04]&8556.8\\ 
1860&198&54&784.9&4.580e-05&[3.441e-05,5.880e-05]&21833.8\\
    \end{tabular}
    \caption{Statistics of lifetimes. For each value of $Re$, $r$ is the number of simulations out of total $n$ to relaminarise before the censoring time $t_r=7800$. The distribution of lifetimes is approximately exponential for $t>t_0$. The maximum likelihood estimator of a mean lifetimes is computed using \eqref{mle} and its confidence interval using \eqref{mle-conf}. The escape rate from the saddle $\kappa=1/\tau$ and its confidence interval are plotted in figure \ref{fig:lt-plot-rev}.   }
    \label{tab:ltt0}
\end{table}
}


%\subsection{Chaotic attractor}
% \begin{figure}
%     \centering
%     % \includegraphics{}
%     \caption{time series lin + log}
%     \label{fig:enter-label}
% \end{figure}
% \begin{figure}
%     \centering
%       \includegraphics[width=0.46\textwidth]{images/jfm-2100-2.pdf}
%     \includegraphics[width=0.49\textwidth]{images/jfm-2100-1.pdf}
  
%     \caption{Caption}
%     \label{fig:enter-label}
% \end{figure}

 The whole situation is reminiscent from the finite probabilities to relaminarise in three-dimensional turbulence in wall-bounded shear flows, notably the case pipe flow, where lifetimes were reported to increase super-exponentially with $Re$ \citep{hof2008repeller}.
For the case of pipe flow, the issue of whether lifetime truly diverge at a finite value of $Re$ or stay finite was \review{solved} by remarking that another competing phenomenon, namely turbulence proliferation, would outweigh turbulence decay above some threshold in $Re$ \citep{avila2011onset} and keep turbulence alive for all times. There is no such equivalent here in axisymmetric rotor-stator flow, for which a scenario where lifetimes are always finite is a possibility \review{below the threshold of linear stability of the base flow $Re_c\approx3000$ \citep{Daube_cf_2002}}. Nevertheless,
% we stress that this debate is essentially academical, %review
the values of $T$ found \emph{in practice} are so large that they usually exceed the observation times allowed for. In other words, the hypothesis that the invariant set underlying the existence of non-trivial dynamics in the forced problem is a deterministic attractor is only here for commodity : mathematically speaking it could possibly be a leaky attractor i.e. a chaotic saddle \citep{lai2011transient}.

\section{Conclusions and outlooks} \label{summary}

%\begin{figure}
%    \centering
%    \includegraphics[width=0.8\textwidth]{images/aps-ctn.png}
%    \caption{Caption}
%    \label{fig:enter-label}
%\end{figure}

The present numerical investigation 
has focused on closed axisymmetric rotor-stator flow as a prototype flow for the study of receptivity. Motivated by the riddle about the origin of the dynamical circular rolls observed in several experiments \citep{Savas_jfm_1987,Gauthier_jfm_1999,Schouveiler_jfm_2001}, both linear and nonlinear receptivity theories were invoked to investigate how the flow responds to an additive forcing of external origin. Several types of forcing were considered, each with its pros and cons. Forcing applied to the bulk of the fluid lends itself well to linear optimal response theory \citep{schmid2014analysis}. If a wide frequency spectrum is forced, pseudo-resonance linked with non-normal effects \citep{trefethen2020spectra} will amplify more the frequencies contained in the large hump in the gain curve in fig. \ref{fig: ressol}. Such frequencies invariably lead to circular rolls inside the Bödewadt layer developing along the stator. Forcing applied through motion of the boundaries is by design not optimal, however it is more realistic in the case of closed flows, and can easily be extended to a nonlinear framework. Increasing the amplitude of the forcing at high enough $Re$ unambiguously leads to a different, more complex nonlinear state which coexists with the simpler forced base flow. Although both states are characterised by circular rolls on the stator, we argue that 
the rolls reported in the experiments of \cite{Schouveiler_jfm_2001} are a linear response to external forcing and not a self-sustained state like those found by \cite{gesla2023subcritical}. This is consistent with the observation that these rolls disappear rapidly should the forcing be suddenly turned off \citep{Lopez_pof_2009,do2010optimal}. Moreover a particularly striking match with the experiment of \cite{Schouveiler_jfm_2001} was found when the forcing features only harmonics of the rotor's main angular frequency, with complex features such as the pairing and merging of vortices well reproduced numerically. The circular rolls observed experimentally can hence be described as noise-sustained, with the subtlety that they are not the response to incoherent noise but rather to a forcing with well-organised frequency content typical of rotating machinery. 

\review{The present study is fully axisymmetric, which implies 
that only axisymmetric modes are involved in the generation of the circular rolls. This excludes other scenarios involving non-axisymmetric modes, for instance triadic instabilities featuring the interaction of a given non-axisymmetric mode with its complex conjugate. A possible extension of our work would allow for non-axisymmetry already at the linear level.}
\review{
% \sout{When the resolvent approach is concerned, instead of costly spatial discretisation in the azimuthal direction, the modal ansatz could be employed.}
The resulting resolvent matrix would then depend on an additional azimuthal wavenumber $m$ and the optimally forced structures would have to be identified in a 2D parameter space $(\omega,m)$. When the time integration of the forced system is concerned a 3D time integration code would be necessary to study the response of the flow to the boundary forcing. While it is theoretically possible that 3D structures are more amplified that the axisymmetric ones, the experimental \citep{Gauthier_jfm_1999} and numerical \citep{do2010optimal} evidence suggest that the response of the forcing is in the form of circular rolls, supporting the  axisymmetric assumption.}

As far as the larger picture is concerned, although linear receptivity is the starting point for transition theory based on an input-output picture, any theory aiming at predicting the onset of bistable dynamics 
needs to be nonlinear.
% should be able to accommodate for nonlinear effects. 
Extending the resolvent formalism to incorporate nonlinear interactions is a possibility \citep{rigas2021nonlinear}.
% \yd{\sout{As is the case in the present study, the optimality idea from classical linear resolvent theory gets lost in the nonlinear framework in favour of other nonlinear selection criteria.}}
A clearer connection between 
% \yd{\sout{the input-output formalism \citep{sipp2013characterization},}}
% \yd{\sout{former}}
theories
% \yd{\sout{with the selection criteria}}
of nonlinear global modes in spatially developing flows \citep{pier2001nonlinear} and the dynamical systems framework would \review{also} be welcome. 
% \lmw{with "however" we seem to denigrate the approach of Rigas et al., do we want to open the debate ? }
% \yd{\sout{A generalisation of the concept of state space, laminar-turbulent boundary, edge states and other finite-amplitude states to forced problems is awaited for, and seems possible given the encouraging results in figure \ref{fig:compLNLAom2}. }}

% \lmw{This last paragraph is  a bit difficult to read.}

\vspace{1cm}

\noindent
{\bf Declaration of Interests :} The authors report no conflict of interest.

{\bf Acknowledgement :} The authors wish to thank anonymous referees for suggestions, in particular the superexponential fit of the lifetimes.

% \bibliography{biblio,localbiblio}
% \bibliography{biblio,localbiblio,jfm-instructions}
% \bibliography{jfm-instructions}
% \bibliographystyle{plain}

\bibliographystyle{jfm}
% \bibliography{biblio,exbiblio_AG,localbiblio_JFM_2024}
\bibliography{external-do_not_modify/biblio,external-do_not_modify/biblio_AG,localbiblio_JFM_2024}
% \bibliography{localbiblio_JFM_2024}

\newpage

% \section*{Yohann's suggestion (18/01/2024)}

% {\bf S1 Introduction}\\
% {\bf S2 Definitions and Numerical Methodology}\\
% {\bf S3 Linear response to (harmonic) forcing}\\
%    a - bulk forcing —> optimal modes via SVD of linear resolvent operator\\
%    b - boundary forcing :  linear resolvent approach vs FFT of probe (linear DNS)\\
%    c - Effect of increasing Re on pseudospectra, gain and $r$-dependency\\
%    d - Focus on low Re=200 (exps), pseudospectra and gain (Q : is linear theory enough?)\\
   
% {\bf S4 Nonlinear receptivity (vs linear) to broadband noise}\\
%    a - synthetic noise \\
%    b - nonlinear (=FFT) vs linear (=resolvent)\\
%    c - Parametric study vs (A0,Re) \\
% {\bf S5 Switching off the forcing}\\
%    a - evidence for rapid collapse at low Re (~ exps) once forcing is switched off\\
%    b - lifetimes at higher $Re$\\
%    c - transition to sustained chaotic state (spectra) at even higher $Re<Re_c$, different between the forced and unforced state depending on parameters\\
% {\bf S6 Conclusions}\\

% Additional:

% Remarks:

% + opt forcing is aligned with shear

% + bulk forcing is like the mean flow correction, response to forcing u'u'

% + first order deviation from Stokes flow will have conncetion to opt forcing / Laurent's idea
% % \newpage

% % \tableofcontents

% \newpage

% \setcounter{page}{1}

\end{document}